\documentclass[12pt]{article}
\usepackage{amsmath}
\usepackage{amsfonts}
\usepackage{amssymb}
\usepackage{srcltx}
\usepackage{graphicx}
\usepackage[title]{appendix}
\def\a{\alpha}

\def\eps{\epsilon}

\def\rf#1{(\ref{eq:#1})}
\def\lab#1{\label{eq:#1}}
\makeatletter
\newcommand{\rd}{\@ifnextchar^{\DIfF}{\DIfF^{}}}
\def\DIfF^#1{%
   \mathop{\mathrm{\mathstrut d}}%
   \nolimits^{#1}\gobblespace}
\def\gobblespace{\futurelet\diffarg\opspace}
\def\opspace{%
   \let\DiffSpace\!%
   \ifx\diffarg(%
   \let\DiffSpace\relax
   \else
   \ifx\diffarg[%
   \let\DiffSpace\relax
   \else
   \ifx\diffarg\{%
   \let\DiffSpace\relax
   \fi\fi\fi\DiffSpace}
\providecommand*{\dder}[3][]{%
\frac{\rd^{#1}#2}{\rd #3^{#1}}}
\providecommand*{\pder}[3][]{%
\frac{\partial^{#1}#2}{\partial #3^{#1}}}
\providecommand*{\iu}%
{\ensuremath{\mathrm{i}\,}}

\begin{document}

\makeatother

\parindent=0cm

\renewcommand{\title}[1]{\vspace{10mm}\noindent{\Large{\bf

#1}}\vspace{8mm}} \newcommand{\authors}[1]{\noindent{\large

#1}\vspace{5mm}} \newcommand{\address}[1]{{ #1\vspace{2mm}}}

\begin{center}

\vskip 3mm

\title{ Solutions of Mixed Painlev\'e P$_{\mathbf{III-V}}$ Model}

 \authors{
V.C.C. Alves$^{1}$,
 H. Aratyn$^{2}$, 
J.F. Gomes$^{1}$
 and A.H. Zimerman$^{1}$
 }

\address{
$^{1}$
Instituto de F\'{\i}sica Te\'{o}rica-UNESP\\
Rua Dr Bento Teobaldo Ferraz 271, Bloco II,\\
01140-070 S\~{a}o Paulo, Brazil

\par \vskip .1in

$^2$ Department of Physics \\
University of Illinois at Chicago\\
845 W. Taylor St.\\
Chicago, Illinois 60607-7059\\
}

\end{center}

\vskip 1mm

 \begin{center}
\textbf{Abstract}
\vskip 3mm
\begin{minipage}{14cm}%

We review  the construction   of the mixed Painlev\'e P$_{III-V}$ system  in terms of  a 4-boson integrable model and  discuss its symmetries.
Such a  mixed system  consist of an  hybrid
differential equation that  for special limits of 
its parameters reduces to  either Painlev\'e P$_{III}$ or P$_{V}$.  

The aim of this paper is to describe solutions of
P$_{III-V}$ model.   In particular,  we determine and classify
rational, power series and transcendental  solutions of 
P$_{III-V}$.  A class of
power series solutions is shown to be convergent in accordance with
the Briot-Bouquet theorem.
Moreover, the P$_{III-V}$  
equations are reduced to Riccati equations and 
solved for special values of parameters.   
The corresponding Riccati solutions can be expressed as
Whittaker functions or alternatively confluent hypergeometric and
Laguerre functions and are given by ratios of polynomials of order
$n$ when the parameter of  P$_{III-V}$ equation is 
quantized by integer $n \in \mathbb{Z}$.

\end{minipage}

\end{center}

\section{Introduction}

 The Painlev\'e equations  were introduced almost  100 years ago  
as second order differential equations that have no movable
critical points other than poles (Painlev\`e property).
Among $50$ second order differential equations that possess the 
Painlev\`e property, $44$ could  either be linearized or 
reduced to ordinary differential equations with solutions expressed 
in terms of known 
transcendental functions. The remaining $6$ equations turned out 
to have solutions 
that are new transcendental functions. We refer
to these $6$  equations as Painlev\'e
P$_{I}$, P$_{II}$, {\ldots} , P$_{VI}$ equations.

The Painlev\`e equations and their solutions are  
ubiquitous in various areas of Mathematical 
Physics. Painlev\'e P$_{III}$ equation emerged 
in studies of Ising and ferromagnetic models \cite{Mccoy1, Mccoy2}.
The transport of ions  under electric field can be studied using a
pair of first order differential equations (Nernst-Planck equations)
\cite{bracken} connected with Painlev\'e P$_{II}$ equation.  
Painlev\'e P$_{V}$ equations  appeared 
in the context of impenetrable Bose gas model  \cite{jimbo}.  
More recently the scattering on two Aharonov-Bohm vortices  was  exactly solved  in terms of  solutions of  the Painlev\'e III equation \cite{Bogomolny}.
 
Ablowitz Ramani and Segur (ARS) showed a connection between the mKdV
integrable model and the Painlev\'e P$_{II}$ equation when 
self-similarity limit was implemented. They suggested 
that such connection could be generalized to other integrable 
models \cite{ablowitz} meaning that
the self-similarity limit would lead for integrable models 
to equations with the Painlev\`e property.
 
More recently,  it was found   \cite{nissimov} that the  
$2n$ boson  integrable model  obtained as particular reductions  
(Drinfeld-Sokolov) of KP integrable models connected 
to  Toda lattice hierarchy gives rise
to higher Painlev\'e equations invariant under 
extended affine Weyl groups.  
In particular, investigation of various Dirac reduction schemes applied to
 the $4$-boson ($n=2$) integrable model with Weyl
symmetry structure $A_4^{(1)}$ led to emergence
of a new mixed P$_{III-V}$ model  \cite{AGZ2016}.
The second order P$_{III-V}$ equation for a canonical
variable $q$ is given by:
\begin{align}\lab{qzz}
		q_{zz}=&-\frac{q_z}{z}+\left(\frac{1}{2q}+\frac{1}{2(q-r_1)}\right)\left(q_z^2-\epsilon_0^2r_0^2z^{4C}\right)-\left(2\alpha_2+\alpha_1+\alpha_3-1\right)\frac{\left(q-r_1\right)qr_0}{z}\nonumber\\
		&+\frac{qr_0^2}{2}(q-r_1)(2q-r_1)+\frac{\alpha_1^2r_1(q-r_1)}{2z^2q}-\frac{\alpha_3^2r_1q}{2z^2(q-r_1)}\nonumber\\
		&+\frac{\epsilon_0r_0z^{-J-2}}{q(q-r_1)}\left((\alpha_1+\alpha_3-J)q^2+qr_1(J-2\alpha_1)+\alpha_1r_1^2\right)-2\epsilon_1r_1z^{J-1}q(q-r_1),
	\end{align}
where $J\equiv-(1+2C)$, $C, r_0 ,r_1, \a_1, \a_2, \a_3, 
\eps_1, \eps_0 $ are all parameters. They will be introduced 
later in the text. For special values of some of these  parameters
P$_{III}$ and P$_{V}$ will emerge from \rf{qzz}.

In reference \cite{AGZ2016} the
Painlev\'e property test was established in the context of equation
\rf{qzz} with arbitrary $\epsilon_i, i=0,1$. The idea of 
combining two different Painlev\'e equations has been explored
in the literature in different settings and a good example  
of such work is \cite{kudryashov}.

In this paper we study the solutions of the mixed  P$_{III-V}$ model
and explore their features that seem to differ from 
the known solutions of the two 
sub-models P$_{III}$ and P$_{V}$.  
  
The paper is organized as follows.
 
In section \ref{sec:intro} we review the derivation of the 
mixed model P$_{III-V}$ equation, its hamiltonian and 
its symmetries as well as the limits,  P$_{V}$ or P$_{III}$ equations,
that will be shown to follow from equation \rf{qzz} 
for special values if its parameters. Apart from serving as a review 
this section also allows us to introduce few fundamental features of the  P$_{III-V}$
model that there not derived previously in \cite{AGZ2016}, namely 
the special form of the Hamiltonian in eq. \rf{hbarzero}, as well as 
automorphisms  given in their explicit forms in this setting in relations \rf{ext-pi} and 
\rf{duality}.

The class of
power series solutions of P$_{III-V}$ model is identified in Section
\ref{sec:BB} and their convergence is established based on the
Briot-Bouquet theorem for a range of its parameters.

In Section \ref{sec:Whittaker} the P$_{III-V}$  
equations are reduced to Riccati equations and 
solved for the three values of the underlying parameter $C$. 
Those solutions reduce to
rational solutions for special values of parameters and it is shown
that they are labeled by $n \in \mathbb{Z}$. In this framework 
we provide examples in subsection \ref{subsection:ratlaguerre} of how
solutions to P$_{III-V}$  equation reduce to solutions of P$_{V}$  and P$_{III}$  
equations when some of the parameters of P$_{III-V}$ model  are
adjusted to zero.

In Section \ref{sec:solutions} we provide
construction and classification of a class of rational solutions 
of  P$_{III-V}$   that are not simultaneously solutions of 
the Riccati equation and are obtained by computer-algebraic calculations.
This construction is presented for cases with one and also two
deformation parameters.  The rational solutions with two
independent parameters are found in the former case, while the 
rational solutions that dependent on three parameters are associated 
with the latter case.
Finally, in Section \ref{sec:discussion} we conclude this paper by
stressing few fundamental features of solutions that possibly reflect
the structure of parameter space of  P$_{III-V}$  model.
Appendix A  lists the alternative solutions to Whittaker solutions of
Riccati equation in terms of the confluent hypergeometric functions and 
Laguerre polynomials that is being used to obtain a large class of
rational solutions.

It would be interesting to establish physical interpretation for a
system of mixed Painlev\'e equations.  This is a long term goal of
our present investigation.

\section{Review of  P$_{\mathbf{III-V}}$ equations and their symmetries}
\label{sec:intro}

In this  section we present an overview of the mixed Painlev\'e 
III - V  model,  its symmetry structure
formed by extended affine Weyl groups acting as
B\"acklund symmetries and its Hamiltonian representation.

The  mixed Painlev\'e 
III - V equations  are defined as  \cite{AGZ2016} :
\begin{align}
z f_{i,\,z} &=f_i  f_{i+2} \big(
 f_{i+1}- f_{i+3} 
 \big)+(-1)^{i} f_i \,\big(\alpha_{1}+\alpha_{3}+C\big)+
\alpha_i\big(f_{i}+f_{i+2}\big)  \lab{big}
\\ 
&-(-1)^{[i/2]} \epsilon_{i+1} \big(f_{i+1}+f_{i+3}\big), \qquad
i=0,1,2,3\, ,
\nonumber
\end{align}
where $f_i=f_{i+4}, \epsilon_i=\epsilon_{i+2}$, the symbol $[i/2]$ is $i/2$, if $i$ is even or $(i+1)/2$, 
if $i$ is odd. $C, \alpha_i,\,i=0,1,2,3$ are constant parameters of
the model. 
Explicitly for $f_1,f_2$ these equations become:
\begin{align}
z f_{1,\, z} &=f_1 f_{3} \big(
 f_{2}- f_{0} \big)-
 f_1 \,\big(\alpha_{1}+\alpha_{3}+C\big)  
 + \alpha_1\big(f_{1}+f_{3}\big)
 + \epsilon_0 \big(f_{0}+f_{2}\big),\lab{f1eq} \\
 z f_{2,\, z} &=f_0 f_{2} \big(
 f_{3}- f_{1} \big)+
 f_2 \,\big(\alpha_{1}+\alpha_{3}+C\big) 
 +
 \alpha_2\big(f_{0}+f_{2}\big)
 + \epsilon_1 \big(f_{1}+f_{3}\big),
 \lab{f2eq}%
 \end{align}
As follows by summing equations  \rf{big} separately 
for even and odd indices the combinations 
$f_0+ f_2$ and $f_1+f_3$ must satisfy constraints:
\begin{equation}
f_1+f_3 =r_1 z^{-C}, \quad
f_0+f_2 = r_0 z^{(C+\Omega) } \, ,
\lab{condcbig}
\end{equation}
where  $r_0,r_1$ are integration constants 
of \rf{big} and $\Omega$ is a constant defined by relation:
\begin{equation}
\Omega= \alpha_0+\alpha_1+\alpha_2+\alpha_3 \ne 0\, .
\lab{omegadef}
\end{equation}
Throughout the paper we will set $\Omega$ to one: $\Omega=1$,
without any loss of generality. Note that the dependence on $z$ on the
right hand sides of equations \rf{condcbig} is identical for
$C=-1/2$. Further, for special cases of $C=0$ and $C=-1$ the sums $f_1+f_3$ or 
$f_0+f_2$, respectively, are constants.

For readers familiar with a conventional symmetric formulation of 
P$_{V}$ equations :
\begin{equation} z f_{i,\,z} =f_i  f_{i+2} \big(  f_{i+1}- f_{i+3} 
 \big)+(-1)^{i} f_i \,\big(\alpha_{1}+\alpha_{3}-\frac12 \big)
+\alpha_i\big(f_{i}+f_{i+2}\big), \;\;i=0,1,2,3\, ,
\lab{P5sym}
\end{equation}
that follow from \rf{big} by inserting $C=-1/2$ and $\epsilon_i=0$, 
and which can be obtained from the Hamiltonian (see e.g. \cite{noumiy,masuda,mok})
\[
\begin{split}
h_0 &= f_0 f_1 f_2 f_3+\frac14 (\alpha_1+2\alpha_2-\alpha_3) f_0 f_1
+ \frac14 (\alpha_1+2\alpha_2+3 \alpha_3) f_1 f_2 \\
&-\frac14 (3\alpha_1+2\alpha_2+\alpha_3) f_2 f_3 
+\frac14 (\alpha_1-2\alpha_2-\alpha_3) f_0 f_3+ 
\frac14 (\alpha_1+\alpha_3)^2
\end{split}
\]
it might be helpful to point out that equations  \rf{big} can be derived from 
the following generalization of the above Hamiltonian:
\begin{equation}
{\bar h}_0= h_0 +\frac{\epsilon_0}{2} (f_0^2-f_2^2)
+\frac{\epsilon_1}{2} (f_1^2-f_3^2) \, .
\lab{hbarzero}
\end{equation}
We refer to $\epsilon_i, i=0,1$ as deformation parameters. 
In the case of vanishing deformation parameters the
equations \rf{big} are invariant under the extended affine Weyl  ${A^{(1)}_{3}}$
symmetry group \cite{Noumiwkb,Noumibk} with parameters $\alpha_0,\alpha_1,\alpha_2$ and $\alpha_3$
entering expressions for ${A^{(1)}_{3}}$
B\"acklund symmetry transformations as follows :
\begin{xalignat}{2}
s_i(\alpha_i)=-\alpha_i, &\quad  s_i(\alpha_k)=\alpha_k+\alpha_i~(k=i 
\pm 1), & s_i(\alpha_k)=\alpha_k~(k \ne i,i \pm 1),\nonumber \\ 
s_i(f_i)=f_i, & \quad s_i(f_k)=f_k \pm \frac{\alpha_i}{f_i}~(k=i \pm 1), 
& s_i(f_k)=f_k ~(k \ne i,i \pm 1),\nonumber \\
\pi(\alpha_k)=\alpha_{k+1}, &\quad  \pi(f_k)=z^{J\,(-1)^{k+1}}f_{k+1}, &
 i,k=0,1,{\ldots} ,3\, ,
 \lab{BT:f}
\end{xalignat}
where for convenience we introduced 
in the last line of equation \rf{BT:f}
the  quantity
\begin{equation}
J=-(1+2C) \, .
\lab{jdef}
\end{equation}
For $C=0,-1/2,-1$, $J$ takes values
$J=-1,0,1$.
Transformations $s_i, \pi, i=0,1,2,3$ from  \rf{BT:f} do not change
the value of the parameter $C$. They satisfy basic $A^{(1)}_3$ relations:
\[
s_k^2=1 , \; (s_is_k)^2=1, \; k\ne i ,i\pm 1, \; 
(s_is_k)^3=1, \; k=i\pm 1,\; \pi^4=1, \; \pi s_i =s_{i+1} 
\pi,
\]
where $i, k=0,1,2,3$.

In general for $\epsilon_i \ne 0, i=0,1$ the mixed Painlev\'e 
equations \rf{big} are only invariant under an 
automorphism $\pi$ generalizing the one defined in 
relation \rf{BT:f} by extending it to act 
 on the parameter space
$\left( \alpha_i, f_i, \epsilon_k, C, r_k \right)$
with $i=0,1,2,3, k=0,1$ via : 
\begin{equation}
\begin{split}
 \pi(\alpha_i)& =\alpha_{i+1}, \quad  \pi(f_i)=f_{i+1}, \; i=0,1,2,3\\
 \pi (\epsilon_0)&= \epsilon_1,  \quad \pi (\epsilon_1)=- \epsilon_0,
 \quad \pi(C)=-1-C, \\
 \pi (r_0)& =r_1, \quad \pi (r_1)=r_0\, .
\end{split}
\lab{ext-pi}
\end{equation} 
We note that the above generalization of an automorphism $\pi$
acts non-trivially on $C$ and $J$ via
$\pi(J)=-J$ with the value $C=-1/2$ or $J=0$ 
fixed under \rf{ext-pi}. 
The constraints \rf{condcbig} 
transform into each other under $\pi$ defined in  \rf{ext-pi}.
One derives from \rf{ext-pi} that  $\pi^2$ : $f_i \to
f_{i+2}, \alpha_i \to \alpha_{i+2}, \epsilon_i \to -\epsilon_i, C \to
C, r_i \to r_i $ will keep the mixed Painlev\'e 
equations \rf{big} invariant without changing the value of $C$.
Also, the above automorphism $\pi$ \rf{ext-pi} keeps the additional $\epsilon$-dependent 
terms in \rf {hbarzero} invariant.

Apart from the obvious case of vanishing deformation parameters
$\epsilon_i=0, i=0,1$  there are also special values of the
parameter $C$ and integration constants $r_i, i=0,1$ for which
equations \rf{big} will have extended symmetry properties.
We will now discuss these special cases. 

First, for the special value of $C=-1/2$ or $J=0$ and the non-zero integration constants
$r_i \ne 0, i=0,1$ and the deformation parameters $\epsilon_i \ne 0, i=0,1$  
the mixed Painlev\'e equations \rf{big} can be shown to remain 
invariant under the above extended
affine Weyl ${A^{(1)}_{3}}$ symmetry group due to the fact that terms with $\epsilon_i$  can
be absorbed in re-definitions of $\alpha_i$'s (without affecting the sum
$\sum_i \alpha_i=1$) and equations \rf{big}
can accordingly be cast in the form of P$_{V}$ equation \rf{P5sym}.

Secondly, in $r_i \to 0$ limits for either $i=0$ or $i=1$ 
the mixed Painlev\'e equations \rf{big} reduce to P$_{III}$  
equations with its own extended affine Weyl symmetry  as have been shown in \cite{AGZ2016}, 
see also equation \rf{qzzr10}  below.

In addition to its invariance under $\pi$ the mixed Painlev\'e III-V 
equations \rf{big} also remain
invariant under another authomorphism 
$\Delta$ :
\begin{equation}
\begin{split}
\Delta (f_{2k-1}) &= (-1)^{1+C} f_{2k}, \;\; \Delta (f_{2k}) =
(-1)^{-C} f_{2k-1},\; k=1,2 \\
\Delta (z)&= - z, \; \Delta (j) = -j ,\; \Delta (r_0) =r_1,\; 
\Delta (r_1) =r_0 ,\\  \Delta (\epsilon_0) =& (-1)^{-J} \epsilon_1,\; 
\Delta (\epsilon_1) = (-1)^{J} \epsilon_0,\\ \Delta (\alpha_i) &=
\alpha_{i+1},\; i=1,3, \; \Delta (\alpha_i) =
\alpha_{i-1}, i=0,2
\end{split}
\lab{duality}
\end{equation}
The repeating
application of $\Delta$ on $f_i$ yields $\Delta^2 (f_i)=(-1)^{(-1)^i 2
C} f_i$ and therefore as long as $C$ is an integer it holds that 
$\Delta^2 =1$ . Also, it holds that $ \left[ \pi^2, \Delta \right]=0$.

When one of the deformations parameters
$\epsilon_0$  or $\epsilon_1$  is zero (with integration constants
$r_i$ being non-zero) the extended automorphism
$\pi$ from relation \rf{ext-pi} is broken although $\pi^2$ remains 
a symmetry. One can show that in  such case $\pi^2$ with $s_{2k}$ or 
$s_{2k+1}$ for $k=0,1$ form
the extended affine Weyl symmetry group 
that keeps the model \rf{big} invariant.
For example setting $ \epsilon_1$ to zero: $ \epsilon_1=0$, 
results in invariance of mixed Painlev\'e  III-V equations \rf{big}
under  B\"acklund  transformations $s_0,s_2$ defined in relation \rf{BT:f}
for any values of $C, r_i \ne 0, i=0,1$ and arbitrary 
$\epsilon_0 \ne 0$. 
In such case $s_0, s_2, \pi^2$ will form
the extended affine Weyl symmetry group $W (s_0,s_2, \pi^2)$:
\begin{equation}
s_0^2=s_2^2=(\pi^2)^2=1,\quad s_0s_2=s_2s_0, \quad 
\pi^2 s_0=s_2 \pi^2 \, ,
\lab{s0s2pi}
\end{equation}
that leaves equations \rf{big} and the value of the parameter $C$
invariant.

For $\epsilon_0 =0$ the B\"acklund 
transformations $s_1,s_3$ defined in relation \rf{BT:f} will leave the 
mixed  Painlev\'e III-V equations
invariant for any values of $C, r_i \ne 0, i=0,1$ and arbitrary
$\epsilon_1\ne 0$ and with $s_1, s_3, \pi^2$ forming
the extended  affine Weyl symmetry group $W (s_1,s_3, \pi^2)$
isomorphic to \rf{s0s2pi}.

Thus in case of only one non-zero deformation parameter $\epsilon_i\ne 0$ one has two
different realizations of the B\"acklund symmetry group
depending on whether one sets $i=0$ or
$i=1$. 
The action of $\pi$ (and $\Delta$) is 
such that it flips between these two different realizations of 
the extended  affine Weyl symmetry groups meaning that $\pi$ takes equation \rf{big} with parameters 
$\left( \alpha_i, f_i, \epsilon_0 \ne 0, \epsilon_1=0, C, r_k \right), 
i=0,1,2,3, k=0,1$ 
and transforms it to the same equation \rf{big} but with the parameters 
$\left( \alpha_i, f_i, \epsilon_0 = 0, \epsilon_1 \ne 0, C, r_k \right), 
i=0,1,2,3, k=0,1$. 

In this context the commutation relations
$\pi s_1 =s_2 \pi $ and $\pi s_3 =s_0 \pi $ etc. 
hold and can be summarized by
\begin{equation}
 \pi s_{2k-1} \pi =s_{2k}, \;\; \pi  s_{2k} \pi =
 s_{2k-1}, \;    \Delta s_{2k-1} \Delta =s_{2k}, \;\; \Delta s_{2k} \Delta =
 s_{2k-1},
 \; k=1,2 \, .
\lab{deltas}
\end{equation}
with $s_{2k-1}$ and $s_{2k}$ on both sides of relations defined for
$\epsilon_0=0$ or $\epsilon_1=0$, respectively.
The relations \rf{deltas} establish the extended $\pi$ as defined in
relation \rf{ext-pi}
as a generalization of the B\"acklund automorphism $\pi$ in the
setting of   mixed Painlev\'e 
P$_{III-V}$ equations. 

Define canonical variables $q,p$ as:
\begin{equation}
q= f_1\, z^C, \; p=-f_2 \, z^{-C},
\lab{qpf}
\end{equation}
then equations \rf{big} can be rewritten %
as two Hamilton equations:
\begin{align}
z q_z &= \pder{{\bar h}_0}{p}=
q \left(q-r_1 \right) \left(2p+r_0 z \right)- 
\left(\alpha_1+\alpha_3 \right) q
+\alpha_1 r_1  +\epsilon_0 r_0 z^{1+2C} \lab{qz}\\
z p_z &=  - \pder{{\bar h}_0}{q}=
p \left(p+ r_0 z \right)
\left(r_1 - 2q \right) + 
(\alpha_1+\alpha_3) p - 
\alpha_2  r_0 z -
\epsilon_1 r_1 z^{-2C}
\lab{pz}
\end{align}
that follow from the Hamiltonian \rf{hbarzero}.

Eliminating $p$ from equation \rf{pz} one obtains
a second order equation \rf{qzz} \cite{AGZ2016}.
Equation \rf{qzz} contains both Painlev\'e V and III equations in special
limits. Setting both deformations parameters $\epsilon_i, i=0,1$
to zero reproduces 
a standard P$_V$ equation
\begin{equation}\lab{p5}
y_{zz} (z)=\left(\frac{1}{y-1}+\frac{1}{2 y}\right)
y_z^2-\frac{y_z}{z}+\frac{(y-1)^2 \left(\alpha 
y+\frac{\beta }{y}\right)}{z^2}+\frac{\gamma y}{z}+\frac{\delta y
(1+y)}{y-1} \, ,
\end{equation}
for 
\[
y = 1 -\frac{r_1}{q}\, ,
\]
with P$_V$ parameters given by :
\begin{equation}
\alpha = \frac{\alpha_1^2}{2},\quad\beta =
\frac{-\alpha_3^2}{2},\quad\gamma
= -\left(-1+\alpha_1+2 \alpha_2+\alpha_3\right) r_0 r_1
,\quad\delta = -\frac{r_0^2 r_1^2}{2}\,.
\lab{PVparameters}
\end{equation}
Taking a limit $r_1 \to 0$ of \rf{qzz} yields
\begin{equation} \begin{split}
q_{zz} & = - \frac1z q_z+  \frac{1}{q}
\left(q_z^2 -\epsilon_0^2 r_0^2 z^{4C}\right)
- (2 \alpha_2+\alpha_1+\alpha_3-1) \frac{q^2 r_0}{z} \\
&+r_0^2 q^3
+ \epsilon_0r_0 z^{-J-2}(\alpha_1+\alpha_3-J) \, .
\end{split}
\lab{qzzr10}
\end{equation}
For $C=0$ or $J=-1$ this equation takes a form of a conventional 
P$_{III}$  equation \cite{okap3}. One can show that this result
extends to any  $C\ne -1$ by a simple rescaling of variables 
while for $C=-1$ (or $J=1$) the model becomes one of $44$ solvable
Painlev\'e equations \cite{paperinpreparation}. Similarly, for $r_0
\to 0$ with $\epsilon_i \ne 0, i=0,1$  equation \rf{qzz}
reduces to P$_{V}$  with the parameter $\delta=0$ in \rf{p5}
for $C\ne 0$, which, as it is well-known \cite{gromak-book}, is fully
equivalent to P$_{III}$ equation. For $C= 0$ the model becomes in the $r_0
\to 0$ limit one of the solvable
Painlev\'e equations  \cite{paperinpreparation}.

In reference \cite{AGZ2016} the
Painlev\'e property test was established in the context of equation
\rf{qzz} with arbitrary $\epsilon_i, i=0,1$. The idea of of
generalizing two different Painlev\'e equations has been explored
in the literature in different settings and good examples  
of such work are \cite{kudryashov}, \cite{rogers}.

In the rest of the paper we focus our attention on the P$_{III-V}$ model with
$\epsilon_0\ne0$ and $\epsilon_1=0$ invariant under 
the extended affine Weyl symmetry group $W (s_0,s_2, \pi^2)$ from
\rf{s0s2pi} and study its solution.
For that reason we will keep in this paper both $r_0 \ne 0$ and 
$r_1 \ne 0$ and, unless explicitly
stated, $C \ne -1/2$ ($J \ne0$).

\section{The Briot-Bouquet System and Power Series Solutions
for  $ \mathbf{J =-(1+2C)\leq 0}$}
\label{sec:BB}
We will derive a class of (convergent) power series 
solutions to equations \rf{qz},\rf{pz} for $J \in \mathbb{Z}$ and $ J \leq 0$
with $\epsilon_1=0$ and $\epsilon_0 \ne 0$. 
The opposite setting of  $\epsilon_1\ne 0$ and 
$\epsilon_0 = 0$ can be described using 
$\pi$
from relation \rf{ext-pi}.

Consider expansion for small $z$ ($J \le 0$) :
\begin{equation} 
q(z) =  \sum_{i=0}^\infty q_i z^i ,\qquad
p(z)= \sum_{i=0}^\infty  p_i z^i = p_0+
p_{1} z^{1} + p_{2} z^{2} +
\cdots  \, .
\lab{pwrsjn}
\end{equation}
There are two such solutions to the system \rf{qz},\rf{pz} that can be 
obtained for $\epsilon_i=0, \;i=0,1$
and their first few terms are
\begin{equation}
p_0=0, \; p_1=\frac{\alpha_2r_0}{-1+\alpha_1+\alpha_3},\;
q_0=\frac{\alpha_1r_1}{\alpha_1+\alpha_3},
\lab{jzerosol1}
\end{equation}
and
\begin{equation}
p_0=\frac{\alpha_1-\alpha_3}{r_1}, \; 
p_1=-\frac{(\alpha_3^2-\alpha_3+\alpha_2\alpha_3-\alpha_2-\alpha_1+\alpha_1^2+\alpha_1\alpha_2)
r_0}{-1+(\alpha_1-\alpha_3)^2},\;
q_0=\frac{\alpha_1r_1}{\alpha_1-\alpha_3},
\lab{jzerosol2}
\end{equation}
Note that the above expression for $p_0$ is finite as we keep $r_1 \ne
0$ as stated above.
For $J <0$  with $\epsilon_0 \ne 0, \epsilon_1=0$ there is a 
solution to \rf{qz},\rf{pz} with the same 
$p_0,p_1,q_0$ as \rf{jzerosol1}
and with higher terms  $p_2,q_1$ given by (for $J=-2,-3,{\ldots} $)
\[
\begin{split}
p_2&=\frac{{\alpha_2}\,{{r_0}}^{2}{r_1}\, \left(
{{\alpha_1}}^{2}-{{\alpha_3}}^{2}+{\alpha_1}\,{\alpha_2}-
{\alpha_2}\,{\alpha_3}-{\alpha_1}+{\alpha_3} \right)}{{\rm Denom}_{p_2}},\\
{\rm Denom}_{p_2} &=-2\,{\alpha_1}-2\,{\alpha_3}+5\,{{\alpha_1}}^{2}+
5\,{{\alpha_3}}^{2}
+10\,{\alpha_1}\,{\alpha_3}-4\,{{\alpha_1}}^{3}-4\,{{\alpha_3}}^{3}+{{\alpha_1}}^{4}+{{\alpha_3}}^{4}\\
&-12\,{{\alpha_1}}^{2}{\alpha_3}-12\,{\alpha_1}\,{{\alpha_3}}^{2}+
4\,{{\alpha_1}}^{3}{\alpha_3}+6\,{{\alpha_1}}^{2}{{\alpha_3}}^{2}+
4\,{\alpha_1}\,{{\alpha_3}}^{3} \, ,\\
q_1&=-\frac{{\alpha_1}\,{{r_1}}^{2}{r_0}\,{\alpha_3}\, \left( -1+{\alpha_3}
+{\alpha_1}+2\,{\alpha_2} \right)}{-{{\alpha_1}}^{2}
-{{\alpha_3}}^{2}-2\,{\alpha_1}\,{\alpha_3}+4\,{{\alpha_1}}^{3}{\alpha_3}+6\,{{\alpha_1}}^{2}{{\alpha_3}}^{2}+4\,{\alpha_1}\,{{\alpha_3}}^{3}+{{\alpha_1}}^{4}
+{{\alpha_3}}^{4}}\\
&=-\frac{{{r_1}}^{2}{r_0}\,{\alpha_1}\,{\alpha_3}\, 
\left( \,{\alpha_2} -\alpha_0\right)}{-{{\alpha_1}}^{2}
-{{\alpha_3}}^{2}-2\,{\alpha_1}\,{\alpha_3}+4\,{{\alpha_1}}^{3}{\alpha_3}+6\,{{\alpha_1}}^{2}{{\alpha_3}}^{2}+4\,{\alpha_1}\,{{\alpha_3}}^{3}+{{\alpha_1}}^{4}
+{{\alpha_3}}^{4}}\, .
\end{split}\]
For $J=-1$ we encounter presence of $\epsilon_0$ in $q_1$. 

To establish convergence of \rf{pwrsjn} we apply 
the Briot-Bouquet  theorem (see e.g.  \cite{gromak}  or 
\cite{gromak-book}) that states that if the system of equations :
\begin{equation}
z u_j^{\prime}= f_j (z,u_1,{\ldots}, u_n), \qquad j=1,{\ldots} ,n\, ,
\lab{BBtheorem}
\end{equation}
with functions $f_j, j = 1, {\ldots}  , n$, that are analytic in some 
neighborhood of the point $z = u_1 = \cdots = u_n = 0$
and satisfy initial conditions 
\begin{equation}
f_j (0, {\ldots}  , 0) = 0  \qquad j=1,{\ldots} ,n \,,
\lab{BBcondition}
\end{equation}
has a formal solution:
\begin{equation}
u_j(z)= \sum_{k=1}^\infty C_k^{(j)} z^k, \qquad C_k^{(j)}
 \in \mathbb{C} \, ,
\lab{ujt}
\end{equation}
then the expansion \rf{ujt} is convergent in a neighborhood of
$z=0$. We will work with a  $n=2$  version of the Briot-Bouquet theorem
with $u_1=u, u_2=v$.

Substituting $q=u+r_1/2$ and $p=-(v+r_0\,z)/2$ or
into 
equations \rf{qz},\rf{pz} with $\epsilon_0\ne0, \epsilon_1=0$
yields :
\begin{equation}
\begin{split}
z u_z &= \frac{r_1(\alpha_1-\alpha_3)}{2} -(\alpha_1+\alpha_3)\, u 
+ v r_1^2/4-u^2v  +\epsilon_0 r_0 z^{(- J) }\, ,\\
z v_z &=  (\alpha_1+\alpha_3 +2\alpha_2-1) r_0\, z - r_0^2 z^2
\,u+(\alpha_1+\alpha_3)\, v +u \,v^2 \,.
\end{split}
\lab{BBqpjpos}
\end{equation}
For $\epsilon_0=0$ we can easily establish connection with 
Painlev\'e V model since in such case \rf{BBqpjpos} 
becomes 
\begin{equation}
\begin{split}
z u_z &= -h -(a+c)\, u +v/4-u^2v\\
z v_z &=  \gamma z + 2 \delta z^2 u+(a+c)\,v +u \,v^2 \, ,
\end{split}
\lab{BBsystem}
\end{equation}
as given in \cite{gromak-filipuk}.  Note that
 $a=\sqrt{2\alpha},c=\sqrt{-2\beta}$ and $\gamma$
are coefficients of  the standard Painlev\'e V equation \rf{p5}, while
$h=(a-c)/2$.

Let $u= \sum_{i=0}^\infty u_i z^i$ and 
$v= \sum_{i=0}^\infty v_i z^i$ 
then the Briot-Bouquet theorem will hold for 
$U= u-u_0= \sum_{i=1}^\infty u_i z^i$ and
$V= v-v_0= \sum_{i=1}^\infty v_i z^i$  
as long the condition
\rf{BBcondition} holds. 
For this condition to hold all constant terms must cancel on the 
right hand side of \rf{BBqpjpos}. For $\epsilon_1=0$ and for 
$\epsilon_0\ne 0$ with $J<0$ 
that amounts to 
\[ 
\begin{split}
0&=\frac{r_1(\alpha_1-\alpha_3)}{2} -(\alpha_1+\alpha_3)\, u_0 +v_0
r_1^2/4-u_0^2\,v_0\, ,\\
0&=(\alpha_1+\alpha_3)\, v_0 +u_0 \,v^2_0 \, ,
\end{split}
\]
since $\epsilon_0 r_0 z^{(- J )} \to 0$ 
for $z \to 0$ and $J<0$.

The above conditions are satisfied automatically for the solution \rf{jzerosol1}
with
\[p_0=0, \; q_0=\frac{\alpha_1r_1}{\alpha_1+\alpha_3}
\; \to \; v_0=0, \; u_0= \frac{\alpha_1r_1}{\alpha_1+\alpha_3}
-\frac{r_1}{2}\, .
\]
Also for the solution \rf{jzerosol2} with
\[
p_0=\frac{\alpha_1-\alpha_3}{r_1}, \; 
q_0=\frac{\alpha_1r_1}{\alpha_1-\alpha_3},\;\to \; 
v_0=-2\frac{\alpha_1-\alpha_3}{r_1},\;
u_0+\frac{\alpha_1r_1}{\alpha_1-\alpha_3}-\frac{r_1}{2}\, ,
\]
the initial conditions are satisfied automatically.

Thus we have used the Briot-Bouquet theorem to establish  the convergence of the power series \rf{pwrsjn}
for negative $J$ and non-zero value of the deformation parameter
$\epsilon_0$. This extends the result that holds for $J=0$ for a
conventional Painlev\'e V model.

\section{Solutions of Riccati Equations}
\label{sec:Whittaker}
The discussion of Riccati equations and its solutions starts
with an observation that equation \rf{pz} is solved
for 
\begin{equation}
p=-r_0 z, \; \epsilon_1=0, \; \alpha_2=1-\alpha_1-\alpha_3 \;\to\;
\alpha_0=0 \,.
\lab{pr0z}
\end{equation}
Note that alternatively we could have chosen $p=0$ and $\alpha_2=0$ as
solutions of equation  \rf{pz} but this configuration can be obtained
from \rf{pr0z} by $\pi^2$ transformation \rf{pisquare}.

Plugging the values \rf{pr0z} in equation \rf{qz} yields
\begin{equation}
\dder{q}{z}= -r_0 q (q-r_1) - \left(\alpha_1+\alpha_3 \right)
\frac{q}{z} + \frac{\alpha_1 r_1}{z} +\epsilon_0 r_0 z^{2C} \, .
\lab{ricattieq}
\end{equation}
For the three values of $J$ given by $J=-1,0,1$ ($C=-1,-1/2,0$) the 
above Riccati equation 
has solutions given in terms of Whittaker functions. 

Substituting $q= (1/(wr_0)) d w/dz$ we obtain a linearized version
of equation \rf{ricattieq} for $w$;
\begin{equation}
z \dder[2]{w}{z}=  \left( {r_0}\,{r_1}z -\alpha_1+\alpha_3 
\right) \dder{w}{z}+\alpha_1 r_0 r_1 w +\epsilon_0 r_0^2
z^{J} w\, ,
\lab{lricattieq}
\end{equation}
which for the value $J=0$ is recognized as Kummer's equation and for
$J=1$ as the extended confluent hypergeometric equation, however
$J=-1$ introduces additional singularity in equation \rf{lricattieq}. 
To systematically obtain  solutions of equation \rf{lricattieq} 
for those three values of $J$ we will apply further
transformations.
Performing an additional substitution and change of a variable:
\[
w(z) = {z}^{-\frac{\alpha_1+\alpha_3}{2}}\, \exp\left(\frac{r_0 r_1
z}{2}\right) f (\tau), \quad \tau = r_0 \,r_1\, z \, ,
\]
transforms the differential equation \rf{lricattieq} into
a differential equation for $f (\tau)$:
\begin{equation}
\dder[2]{f}{\tau}+ \left( - \frac{1}{4} - 
\frac{\epsilon_0 r_0^2\tau^{2C}}{(r_0
r_1)^{2+2C}}+ \frac{\alpha_3-\alpha_1}{2} \frac{1}{\tau}+
\left(\frac{1}{4} -\left(\frac{\alpha_1+\alpha_3-1}{2}\right)^2 
\right)\frac{1}{\tau^2}\right) f(\tau)=0\, ,
\lab{confhyper}
\end{equation}
which  for the three values of $2C = -1, -2, 0$ or $J=-1,0,1$ 
can be cast as a second order differential Whittaker's equation :
\begin{equation}
f^{\prime\prime}(x) + \left( - \frac{1}{4}
+ \frac{a}{x}+
\left(\frac{1}{4} -b^2 \right)\frac{1}{x^2}\right) f(x)=0\, ,
\lab{confhyper1}
\end{equation}
with a solution:
\[
f (x) = c_1 {\rm WhittakerW} (a,b,x) +c_2 {\rm WhittakerM}  (a,b,x) \, ,
\]
given in terms of the Whittaker functions 
${\rm WhittakerW}$ and ${\rm WhittakerM}$. See Appendix A for
alternative solutions to equation \rf{confhyper1} in terms of confluent
hypergeometric functions and generalized
Laguerre polynomials. The above derivation of solutions to the Riccati
equation \rf{lricattieq} by reformulating this equation 
as Whittaker equation suggests that solutions of \rf{lricattieq} only 
exist for $J=-1,0,1$. This observation is further strengthened by the fact that
neither Maple or Mathematica was able to find solutions to equation
\rf{lricattieq} for other cases
of $J$.

Note that for $C=-1, -1/2$ it holds that $x=\tau$ while for $C=0$ 
rewriting equation \rf{confhyper} as
\rf{confhyper1} requires an additional change of a variable from 
$\tau$ to $x \ne \tau$.
In summary we obtain the following solutions 
of equation \rf{lricattieq}:
\begin{align}
w_{C=0}&= {z}^{-\frac{\alpha_1+\alpha_3}{2}}\, \exp\left(\frac{r_0 r_1
z}{2}\right) \left( c_1
{\rm WhittakerW} 
\left(  \frac{r_1  (\alpha_3-\alpha_1)}{2 \sqrt{r_1^2+4 \epsilon_0}},  
\frac{\alpha_1+\alpha_3-1}{2},r_0 \sqrt{r_1^2+4 \epsilon_0}z 
\right) \right.\nonumber\\
&+\left. c_2 {\rm WhittakerM} 
\left(  \frac{r_1  (\alpha_3-\alpha_1)}{2 \sqrt{r_1^2+4 \epsilon_0}},  
\frac{\alpha_1+\alpha_3-1}{2},r_0 \sqrt{r_1^2+4 \epsilon_0}z 
\right) \right)\, , \lab{wc0}\\
w_{C=-1}&= {z}^{-\frac{\alpha_1+\alpha_3}{2}}\, \exp\left(\frac{r_0 r_1
z}{2}\right) \left( c_1 {\rm WhittakerW} 
\left(  \frac{\alpha_3-\alpha_1}{2},  
\frac12\, \sqrt{(\alpha_1+\alpha_3-1)^2+4r_0^2\epsilon_0},r_0 
r_1z  \right) \right.\nonumber\\
&+\left. c_2  {\rm WhittakerM} 
\left(  \frac{\alpha_3-\alpha_1}{2},  
\frac12\, \sqrt{(\alpha_1+\alpha_3-1)^2+4r_0^2\epsilon_0},r_0 
r_1z  \right) \right)\, , \lab{wcm1}\\
w_{C=-1/2}&= {z}^{-\frac{\alpha_1+\alpha_3}{2}}\, \exp\left(\frac{r_0 r_1
z}{2}\right) \left( c_1
{\rm WhittakerW} 
\left(  \frac{\alpha_3-\alpha_1-2 \epsilon_0r_0/r_1}{2},  
\frac{\alpha_1+\alpha_3-1}{2},r_0 
r_1z  \right) \right.\nonumber\\
&+\left. c_2  {\rm WhittakerM} \left(  \frac{\alpha_3-\alpha_1-2 \epsilon_0r_0/r_1}{2},  
\frac{\alpha_1+\alpha_3-1}{2},r_0 
r_1z  \right) \right) \, .\lab{wcmhalf}
\end{align}
We first note that all three solutions \rf{wc0}-\rf{wcmhalf} for 
$\epsilon_0=0$ coincide with the P$_{V}$ solution derived from Riccati
equation for $\epsilon_0=0$ as expected from construction of the 
P$_{III-V}$ model. For application of Whittaker functions to 
construct solutions of Painlev\'e V equation see e.g.
\cite{shimomura}.

We will now address a question whether there exist algebraic relations
between  the above solutions \rf{wc0}-\rf{wcmhalf} with different values of 
the parameters. We are only interested in keeping
$\epsilon_0 \ne 0$ in this discussion as otherwise the underlying Riccati equation becomes simply
the Riccati equation of P$_{V}$ equation. 
We first observe that the expression for $w_{C=-1/2} $ 
will agree with the solution of \rf{lricattieq} for $\epsilon_0=0$
when we replace $\alpha_3 - \epsilon_0 r_0 /r_1$ by $\alpha_3$ and
$ \alpha_1 + \epsilon_0 r_0 /r_1$ by $\alpha_1$. 
Thus by simple redefinition of the parameters $\alpha_1,\alpha_3$ in a
way that does not change the condition \rf{omegadef} we are able to 
obtain from \rf{wcmhalf}
a solution to P$_{V}$ equation. This follows from the fact 
that, as explained in the introductory
Section \ref{sec:intro},  P$_{III-V}$ model with $J=0$ ($C=-1/2$) agrees
effectively with the model with zero deformation parameters. 
However similar attempts to find algebraic 
relations between expressions \rf{wc0}, \rf{wcm1} and \rf{wcmhalf}
fail as long as $\epsilon_0 \ne 0$. For example rewriting the second
argument of \rf{wcm1} in the same form as the second argument of
\rf{wcmhalf} :
$(\alpha_1^\prime+\alpha_3^\prime-1)/2$ with some new parameters
$\alpha_1^\prime,\alpha_3^\prime$ introduces an extra
dependence on $z$ from the term ${z}^{-\frac{\alpha_1+\alpha_3}{2}}$ in \rf{wcm1}.
Also, there is no B\"acklund transformation that would relate these
three solutions as neither $s_0, s_2$ or $\pi^2$ are able to change
the value of the parameter $J$ labeling these three solutions.

Due to $\pi^2$ symmetry :
\begin{equation}
 \pi^2(q)=r_1-q, \; \pi^2(p)=-p - r_0 z, \; \pi^2(\alpha_i)=
\alpha_{i+2}, \; \pi^2(\epsilon_i)=-\epsilon_i, \, i=0,1
\lab{pisquare}
\end{equation}
of equations \rf{qz},\rf{pz} the solution 
with $p=-r_0 z, \; \epsilon_1=0, \;
\alpha_0=1-\alpha_1-\alpha_3-\alpha_2=0$ is transformed
by $\pi^2$ to a solution with 
$p=0$ and $\alpha_2=0, \epsilon_1=0$ and with the same
value of $C$

The above two families of solutions give rise via the $\pi$ 
operation to two other classes of solutions; one with 
$q=r_0$ and $\alpha_3=0, \epsilon_0=0$
 and 
$\epsilon_1\ne 0$ and the other one with 
$q=0$ and $\alpha_1=0, \epsilon_0=0$. 
Recall that $\pi$ maps $J \to -J$ or $C \to -(1-C)$ and thus the 
solution with $C=0$ is mapped to one with $C=-1$ and vice versa, 
while the $C=-1/2$ 
solution is mapped to another solution with an unchanged value of $C$.

\subsection{Rational Solutions from Whittaker functions}
\label{subsection:ratwhittaker}
The Whittaker functions take simpler expressions when their 
first two arguments differ by $\pm 1/2$: 
\begin{align}
{\rm WhittakerW} (a,a-1/2,z)&={\rm WhittakerM} (a,a-1/2,z)
\nonumber\\
&=
{\rm WhittakerW} (a,-a+1/2,z) 
=z^a
\exp(-z/2),  
\lab{Whittid1}
\end{align}
and
\begin{equation}
{\rm WhittakerM} (a,-a-1/2,z)= z^{-a}\exp(z/2) \, .
\lab{Whittid2}
\end{equation}
 
For $C=0$ and for special values of $\epsilon_0$ such that :
\[ \frac{r_1  (\alpha_3-\alpha_1)}{ \sqrt{r_1^2+4 \epsilon_0}}=  
\alpha_1+\alpha_3, \qquad 
\frac{r_1  (\alpha_3-\alpha_1)}{2 \sqrt{r_1^2+4 \epsilon_0}}=
-\frac{\alpha_1+\alpha_3-2}{2}\, .
\]
Due to Whittaker relations \rf{Whittid1} the expressions
\rf{wc0} and \rf{wcm1} simplify considerably and yield
rational solutions. For an ansatz 
$p=-r_0 z, \; \epsilon_1=0, \;
\alpha_0=1-\alpha_1-\alpha_3-\alpha_2=0$ we find
for $C=0, J=-1$ from \rf{Whittid1}:
\begin{equation} \begin{split}
q(z) &=\frac{\alpha_1 r_1}{\alpha_1+\alpha_3}, \quad \qquad \qquad 
\epsilon_0
=-\frac{\alpha_1\alpha_3 r_1^2}{(\alpha_1+\alpha_3)^2}\\
q(z)&=
\frac{r_1(\alpha_3-1)}{\alpha_1+\alpha_3-2}+\frac{1-\alpha_1-\alpha_3}{r_0
z}, \quad  \epsilon_0
=-\frac{(\alpha_1-1)(\alpha_3 -1)r_1^2}{(2-\alpha_1-\alpha_3)^2}\, .
\end{split}
\lab{qratC=0}
\end{equation}
For $C=-1, J=1$ similar considerations based on Whittaker relations
\rf{Whittid1} yield :
\begin{equation} \begin{split}
q(z) &=-\frac{\alpha_1 }{r_0z}, \quad \epsilon_0
=\frac{\alpha_1(1-\alpha_3)}{r_0^2}\\
q(z)&=
r_1 -\frac{\alpha_3}{r_0 z}
, \quad  \epsilon_0
=\frac{(1-\alpha_1)\alpha_3 }{r_0^2}\, ,
\end{split}
\lab{qratC=m1}
\end{equation}
where the first formula have been derived from \rf{Whittid1} and the
second formula from \rf{Whittid2}. Thus the case of $C=-1$ contains both
proper rational solutions and solutions that are proper up to a constant.

Other similar solutions can be obtained from the above solutions 
by actions of $\pi^2$, and $s_i, i=0,2$.
For example the first of equations in \rf{qratC=0} 
transforms under $\pi^2$ to another solution with $J=-1$:
\begin{equation}
p=0, \quad q(z) =\frac{\alpha_1 r_1}{\alpha_1+\alpha_3}, \quad 
\alpha_2=0, \quad 
\epsilon_0
=\frac{\alpha_1\alpha_3 r_1^2}{(\alpha_1+\alpha_3)^2}\,.
\lab{2algebraic}
\end{equation}
All three transformations $s_0, s_2$ and $\pi^2$ 
preserve the constraint $\alpha_1=\alpha_3$. In all three cases 
new solutions have the same dependence on $z$ and can be  obtained 
through internal re-definitions of $\alpha_i$-parameters generated by these
transformations.
\subsection{Rational Solutions in terms of Laguerre polynomials}
\label{subsection:ratlaguerre}

There is a class of rational solutions that one can obtain from
equation \rf{wc0laguerre}
in the Appendix \ref{sec:alternativeRiccati} for $C=0$ with $c_1=0$ and with the condition
\begin{equation}
\sqrt{r_1^2+4 \epsilon_0} =
\frac{r_1(\alpha_3-\alpha_1)}{2n+\alpha_1+\alpha_3} , \qquad n \ge 0,\;
\text{integer}\, ,
\lab{nepsilon0C0}
\end{equation}
which could be  rewritten as
\begin{equation}
\epsilon_0=-r_1^2 \frac{n^2+\alpha_1\alpha_3+n(\alpha_1+\alpha_3)}{(2n
+\alpha_1+\alpha_3)^2}, %
\lab{nepsilon0C0}
\end{equation}
to show that $\epsilon_0$ is being ``quantized'' in terms of 
integer $n$ under the condition \rf{nepsilon0C0} and fully expressed 
by $n$ and 
$\alpha_1,\alpha_3$.

The condition \rf{nepsilon0C0} ensures that the first argument of
solution  \rf{wc0laguerre} is a non-negative integer $n$ and the equation \rf{wc0laguerre}
is expressed in terms of the associated Laguerre polynomial 
${\text L} [ n, \alpha_1 +\alpha_3-1, z r_0 \sqrt{r_1^2+4 \epsilon_0}]$.
Substituting such solution into $q= (1/(wr_0)) d w/dz$ one obtains
\begin{equation}
\begin{split}
q_{C=0,n} (z)&= \frac{r_1 (n+\alpha_1)}{2n+\alpha_1+\alpha_3}\\
&- \frac{r_1
(\alpha_3-\alpha_1)}{2n+\alpha_1+\alpha_3}
\frac{{\text L} [n-1,\alpha_1+\alpha_3, z r_0r_1
(\alpha_3-\alpha_1)/(2n+\alpha_1+\alpha_3)]}{{\text L}
[n,\alpha_1+\alpha_3-1, z r_0r_1
(\alpha_3-\alpha_1)/(2n+\alpha_1+\alpha_3)]} \, ,
\lab{qC0laguerre}
\end{split}
\end{equation}
for $n >0$ and integer. For $n=1$ one obtains:
\begin{equation}
\begin{split}
q_{C=0,n=1} (z)&=  \frac{ r_1(1+\alpha_1)}{2+\alpha_1+\alpha_3}+ \frac{r_1
(\alpha_3-\alpha_1)}{z r_0 r_1 (\alpha_3-\alpha_1) -
(\alpha_3+\alpha_1)(2+\alpha_1+\alpha_3)},\\
\epsilon_0&=-r_1^2
\frac{1+\alpha_1\alpha_3+(\alpha_1+\alpha_3)}{(2
+\alpha_1+\alpha_3)^2}\, .
\end{split}
\lab{qC0n1}
\end{equation}
For special values of $\alpha_1=-1$ or $\alpha_3=-1$ in the above
formula the deformation parameter $\epsilon_0$ vanishes and subseqently $q_{C=0,n=1} (z)$
solves Painlev\'e V 
equation. We 
find that for $\alpha_1=-1$, $y(z) = 1 - r_1/q_{C=0,n=1}
(z)=-r_0r_1z+\alpha_3$ solves P$_{V}$ equation \rf{p5} with the 
parameters
\rf{PVparameters} given by $\alpha=1/2,
\beta=-\alpha_3^2/2,\gamma=\alpha_3-2, \delta=-r_1^2r_0^2/2$.
Setting  $\alpha_3=-1$ results in $y(z)=1/(r_0r_1z+\alpha_1)$
that solves P$_{V}$ equation \rf{p5} with  $\alpha=\alpha_1^2/2,
\beta=-1/2,\gamma=\alpha_1-2, \delta=-r_1^2r_0^2/2$. 
Setting $r_1=0$ in the formula \rf{qC0n1} yields in agreement with
equation \rf{qzzr10} a relatively
simple solution $q=1/(r_0z)$ of P$_{III}$ equation:
\[y_{zz}  = - \frac1z y_z+  \frac{y_z^2 }{y}
+{\cal A} \frac{y^2 }{z} +{\cal C} y^3
+ \frac{{\cal B}}{ z}+
\frac{{\cal D}}{y}\]
with ${\cal A}=\alpha_1+\alpha_3-1, {\cal C}=r_0^2, {\cal B}=0,
{\cal D}=0$.

Generally the expression \rf{qC0laguerre} can be written as a ratio of
polynomials of order $n$.

The above formula extends to $n=0$ due to ${\text L}
[0,\lambda, x]=1$ and reproduces for this value of $n$ the first of
equations in \rf{qratC=0}.  Inserting $n=-1$ into \rf{nepsilon0C0} and
into equation \rf{wc0laguerre} with $c_2=0$ reproduces the second of
equations in \rf{qratC=0}. Remarkably both expressions \rf{nepsilon0C0} and
equation \rf{wc0laguerre} with $c_2=0$  allow for an extension to 
negative values of $n$. For example for $n=-2$ one gets 
\begin{equation}
\begin{split}
q_{C=0,n=-2} (z)&= \frac{ r_1(-2+\alpha_1)}{-4+\alpha_1+\alpha_3}
-\frac{-2+\alpha_1+\alpha_3}{zr_0}\\
&-\frac{(-2+\alpha_1+\alpha_3)(-4+\alpha_1+\alpha_3)}
{z
r_0(8+\alpha_1^2+\alpha_3(-6+\alpha_3-zr_0r_1)+\alpha_1(-6+2\alpha_3+zr_0r_1))}\\
\epsilon_0&=-r_1^2 \frac{4+\alpha_1\alpha_3-2(\alpha_1+\alpha_3)}{(-4
+\alpha_1+\alpha_3)^2},
\end{split}
\lab{qC0nm2}
\end{equation}
Also in this case setting $\alpha_1=-2$ or $\alpha_3=-2$ to ensure 
that the deformation parameter $\epsilon_0$ vanishes yields 
solutions to P$_{V}$ equation that this time also depend on a second 
order polynomial in $z$. Repeating this contruction for higher values
of $n$ would produce  P$_{V}$ solutions in terms of polynomials 
of higher order.

See subsection \ref{subsec:C0} for generalization of the above 
solutions to other values of integer $n$.

Correspondingly, for $C=-1$ and the Appendix equation \rf{wcmonelaguerre}
by setting 
\begin{equation}
\epsilon_0 = \frac{1}{r_0^2} (n +\alpha_1)(n+1-\alpha_3) , \qquad n \ge 0,\;
\text{integer}
\lab{nepsilon0Cm1}
\end{equation}
one ensures that the first argument of the generalized Laguerre
function is a non-negative integer and the corresponding rational $q$
is given by the formula :
\begin{equation}
q_{C=-1,n} = -\frac{n+\alpha_1}{r_0z}-r_1 \frac{{\text L} [n-1,-2n -\alpha_1+\alpha_3, z r_0r_1
]}{{\text L}
[n,-2n-\alpha_1+\alpha_3-1, z r_0r_1
]}  , \qquad n \ge 0,\;
\text{integer}\, ,
\lab{qCm1laguerre}
\end{equation}
which  for $n=1$ is given by
\[
q_{C=-1, n=1} (z) = -\frac{1+\alpha_1}{r_0z}
+\frac{r_1}{r_0r_1z+2+\alpha_1-\alpha_3}, \qquad 
\epsilon_0 = \frac{1}{r_0^2} (\alpha_1+1)(2-\alpha_3))\, .
\]
The expression \rf{qCm1laguerre} can be cast as a ratio of
polynomials of order $n$.
Also, in this case the formula \rf{nepsilon0Cm1} extends beyond values
of $n$. 
Setting $n=0,n=-1$ we obtain respectively the first and second formula in  
equation \rf{qratC=m1}.
See subsection \ref{subsec:Cm1} for generalization of the above 
solutions to other values of integer $n$.

\section{Construction and classification of solutions 
by  by computer-algebraic  calculations}
\label{sec:solutions}

\subsection{Necessary conditions for rational solutions \'a
la Kitaev, Law, and McLeod}
\label{subsection:kitaev}

Before we present a method to derive a class of rational solutions
based on computer calculations let recall the Proposition 2.2 in 
\cite{kitaev} that lists necessary conditions for
rational solutions of Painlev\'e V equation.
In this subsection we give result of a similar approach 
applied to equation \rf{qzz} for $J= \pm 1$ cases. 
For large $z$ we investigate expressions of the type:
$q=K_0 + a/z+ b/z^2+ c/z^3  + 0(z^{-4})$.
with $K_0$ being a constant.

For the case of $C=0, J=-1$ we find  that  solutions exist
for two values of $K_0$:

(1) $K_0=r_1/2$  with 
\[a=r_1^2\frac{r_1^2 (1-\alpha_1-\alpha_3-2\alpha_2)
+4 \epsilon_0(\alpha_1+\alpha_3-1)}{r_0(-r_1^4+16 \epsilon_0^2)}
\]
\[
\begin{split}
b&= \frac{-2r_1^3}{r_0^2(-r_1^4+16 \epsilon_0^2)}
(r_1^4(\alpha_1^2-\alpha_3^2)\\
&+8
\epsilon_0r_1^2(-\alpha_1^2+\alpha_3^2+\alpha_1
-\alpha_3-2\alpha_1\alpha_2+2\alpha_2\alpha_3)+16\epsilon_0^2
(\alpha_1^2-\alpha_3^2-2\alpha_1+2\alpha_3))
\end{split}
\]
(2) $K_0= \frac{1}{2} r_1  
\pm \frac{1}{2} \sqrt{r_1^2 \pm 4 \epsilon_0}$.
For instance for
\[ q=\frac{1}{2} r_1  + \frac{1}{2} \sqrt{r_1^2 - 4 \epsilon_0}
 + a/z+ b/z^2+ c/z^3  + 0(z^{-4})
\]
one finds:
\[ a= \frac{1}{2 r_0(-r_1^2 + 4 \epsilon_0)} (-r_1^2(\alpha_1+\alpha_3)+
r_1\sqrt{r_1^2 - 4
\epsilon_0}(\alpha_1-\alpha_3)+4\epsilon_0(\alpha_1+\alpha_2+\alpha_3))
\]
etc.
 
For the case $C=-1, J=1$ we find that solutions exist for the following
values of $K_0$:
\begin{enumerate}
\item 
$K_0=0$ (see e.g. the first of equations 
\rf{qratC=m1})
\[ a= \pm \frac{\alpha_1}{r_0}, ; 
b= \pm \frac{ -\alpha_1(\alpha_3+2 \alpha_2-1)+ \epsilon_0
r_0^2}{r_1r_0^2}
\]
\item $K_0=r_1/2$ (see e.g. solution 
\rf{single2para}) exists with
\[ a= \frac{\alpha_1+\alpha_3+ 2 \alpha_2 -1}{r_0} , \;
b= \frac{2(-\alpha_1^2+\alpha_3^2)}{r_1r_0^2}
\]
Remarkably, the series truncates for the choice $\alpha_1=\alpha_3$ with 
$b$ and higher terms vanishing as well.
\item
$K_0=r_1$ (see e.g. the second of equations 
\rf{qratC=m1}) exists with
\[ a= \pm \frac{\alpha_3}{r_0}, ; 
b_{+}= - \frac{\alpha_3(1-\alpha_1-2 \alpha_2)+ \epsilon_0
r_0^2}{r_1r_0^2}, \;
b_{-}=\frac{\alpha_3(1-\alpha_1-2 \alpha_3-2\alpha_2)+ \epsilon_0
r_0^2}{r_1r_0^2}
\]
For $a=-\alpha_3/r_0$ and $\alpha_2=1-\alpha_1-\alpha_3$ 
the above solution agrees with the second of expressions in
\rf{qratC=m1}.
\end{enumerate}
We will see in what follows that the class of solutions obtained below will agree
with the above results.

\subsection{A method for deriving a class of rational solutions for
$\mathbf{q(z)}$.}

Equations  \rf{qz} and \rf{pz} with  $\epsilon_0\ne 0, \epsilon_1 =0$
are invariant under B\"acklund transformations $s_0,s_2$ that
transform $q$ but leave $p$ unchanged. It is therefore natural to
classify solutions of equations  \rf{qz} and \rf{pz} in terms of $p(z)$.

Our method is based on an assumption of $p(z)$ being 
a finite Laurent series $p(z) = \sum_{k=-n_1}^{n_2} p_k z^k $ with finite integers
$n_1,n_2$. 

One can verify from equations \rf{qz} and \rf{pz} that if we start with
$p(z) = \sum_{k=-n_1}^{n_2} p_k z^k $  then the consistency of \rf{qz}
and \rf{pz} requires that $p(z)$ is limited to expression:  
\begin{equation}
p(z)=p_{-J} z^{-J} +p_0+p_1 z \, .
\lab{pjansatz}
\end{equation}
with only three allowed terms containing $z$ of power $0$, $1$ 
and $-J$. $p_{-J} z^{-J} $  is a separate term from $p_0$
and $p_1z$ as long as $J \ne 0, -1$.  It is possible to verify that
the expression \rf{pjansatz} with $J \ne 0, -1$ and $p_{-J} \ne 0$ and $p_0
=0$  requires that $q(z)$ is either equal to $r_1/2$ or a B\"acklund
transform of $q(z) =r_1/2$. 
We therefore start by considering the special case of $q(z) =r_1/2$.

\subsection{The case of  ${\mathbf q =} {\mathbf {\frac{r_1}{2}} }$.}

Inserting $q(z)=\frac{r_1}{2}$ into \rf{qz} we obtain:
\begin{equation}
p(z)=\frac{ (\alpha _1- \alpha _3 )}{ r_1} -\frac{z\, r_0
\left(r_1^2-4 z^{2 C} \epsilon _0\right)}{2 r_1^2} \,.
\lab{pfromqr1over2}
\end{equation}
Plugging this expression  together with $q(z)=\frac{r_1}{2}$
into \rf{pz} we arrive at the equation :
\begin{equation}
0= 2 r_1 (\alpha _1^2- \alpha _3^2)+z \, r_0 r_1^2 \left(1-\alpha_1-2 \alpha_2- \alpha_3\right)+
z^{1+2 C} r_0\epsilon _0 \left(-4 -8 C +4 \alpha _1 +4 \alpha _3
\right) \,.
\lab{psolj}
\end{equation}

When studying expression \rf{psolj} we need to consider separately the three cases of the parameter $C$ 
taking  values $C=0$ or
$C=-1/2$  or the remaining case of $C\ne 0,\,C\ne -1/2$. 

Solving for $\alpha_1,\,\alpha_2,\,\alpha_3$ for all three cases we
obtain (for $r_0 \ne 0, r_1 \ne0$) :
\begin{align}
C&=0\; : \; \alpha_3 = -\alpha_1,\; \alpha _2= \frac{1}{2}-\frac{2\epsilon _0}{r_1^2}
\;\, \text{or}\;\, \alpha _1= \alpha _3,\;\alpha _2= -\frac{\left(-1+2 \alpha _3\right) 
\left(r_1^2-4 \epsilon _0\right)}{2 r_1^2} \lab{Cequal0sols}\\
C&=-\frac{1}{2} : \; \alpha _3= -\alpha _1,\;\alpha _2= \frac{1}{2}
\;\,\text{or}\;\, \alpha _3= \alpha _1+\frac{2 r_0 \epsilon _0}{r_1},\;\alpha _2= \frac{1}{2}-\alpha _1-
\frac{r_0 \epsilon _0}{r_1}
\lab{C=monehalfsols}\\
C&= \text{other}\; : \; \alpha _1 = \frac{1}{2}+C,\; \alpha _2=
-C,\;\alpha _3= \frac{1}{2}+C. \lab{Cothersols}
\end{align}
We notice that all of the above cases with $\alpha_3=\alpha_1$ 
will have a
common solution  :
\begin{equation}
p(z)=\frac{2 z^{1+2 C} r_0 \epsilon _0}{r_1^2}-\frac{z r_0}{2} \, ,
\lab{pgeneral}
\end{equation}
with $p(z)$ derived from eq. \rf{pfromqr1over2}.

For $\alpha_1=-\alpha_3 \ne 0$ we obtain
\begin{equation}
p(z)= \frac{2\alpha_1}{r_1} -\frac{z\, r_0 \left(r_1^2-4 z^{2 C}
\epsilon _0\right)}{2 r_1^2}, \qquad C=0, -1/2 \, .
\lab{pc0mhalf}
\end{equation}
We notice that for the special case of $\alpha_1=0$ the above equation agrees with the formula
\rf{pgeneral}.
Finally for $\alpha _3= \alpha _1+{2 r_0 \epsilon _0}/{r_1}$ and
$C=-1/2$ we obtain $p=-zr_0/2$.

We note that this class of solutions also includes expressions for
 rational functions $q(z)$ obtained from acting on $q=r_1/2$ by  B\"acklund
transformations $s_0$ and $s_2$. 
For example for $C=-1$ we obtain multiplet of solutions by first substituting
$C=-1$ into $\alpha_1 = \frac{1}{2}+C,\; \alpha_2=
-C,\;\alpha_3= \frac{1}{2}+C$ and then acting on $q=r_1/2$ by 
$s_2,s_0, s_0 s_2=s_2s_0$, obtaining respectively :
\begin{align*}
\alpha_{C=-1}\equiv&(\alpha_1=\alpha_3= -\frac{1}{2},\alpha_2= 1)
,&q_0 (z)\equiv&\frac{r_1}{2}\\
s_2(\alpha_{C=-1})=&(\alpha_1=\alpha_3= \frac{1}{2},\alpha_2= -1),&
s_2 (q_0) (z)=&\frac{r_1 \left(-4 z r_1+z^2 r_0 r_1^2-4 r_0 \epsilon _0\right)}{2 r_0 \left(z^2 r_1^2-4 \epsilon _0\right)}\\
s_0(\alpha_{C=-1})=&(\alpha_1=\alpha_3= \frac{1}{2},\alpha_2= 1),
&s_0(q_0) (z)=&\frac{4 z r_1^2+z^2 r_0 r_1^3+4 r_0 r_1 \epsilon _0}{2 z^2 r_0 r_1^2+8 r_0 \epsilon _0}\\
s_0s_2(\alpha_{C=-1})=&(\alpha_1=\alpha_3= \frac{3}{2},\alpha _2= -1),
&s_0s_2 (q_0) (z)=&\frac{r_1 \left(-32 z r_1 \epsilon _0+r_0 \left(z^4 r_1^4-16 \epsilon _0^2\right)\right)}{2 r_0 \left(z^4 r_1^4-16 \epsilon _0^2\right)}
\end{align*}
The above solutions share the same $p(z)$:
\[
p(z)=r_0 \left(-\frac{z}{2}+\frac{2 \epsilon _0}{z r_1^2}\right) \, 
\]
and form a class of  one-parameter solutions with a free parameter $\epsilon_0$.

\subsection{A class of two-parameter solutions}
Starting with an the ansatz :
\begin{equation}
p(z)=p_0+p_1 z \, .
\lab{pansatz}
\end{equation}
(with the term $p_J z^{-J}$ in \rf{pjansatz} removed)
gives rise to a wider class of rational solutions for $q(z)$. We will 
describe here a class of $2$-parameter solutions corresponding to the
ansatz \rf{pansatz} that exists for special values of $C= \pm 1/2, 0,-1$.

Plugging expression \rf{pansatz}
into equations \rf{qz} and \rf{pz} yields 
\begin{equation}
q(z)=  \frac{1}{2} \left( r_1+\frac{-p_1 z+p(z)
(\alpha_1+\alpha_3)-\alpha_2 r_0z}{p(z) (p(z)+r_0 z)} \right) \, .
\lab{qansatz}
\end{equation}
Note that the Riccati cases of $p(z)=0$ and
$p(z)=-r_0 z$,  that were already discussed separately, lead to 
divergences in the above expression for $q(z)$ and
therefore can not be obtained by this method.

Below we obtain closed expressions for special values of $C= -1,0,
+1/2$.

For $C=-1/2$ as expected the case reduces to the standard Painlev\'e V
system since the term $\epsilon_0 r_0 z^{1+2C}$ becomes a
constant and the epsilon terms can be absorbed by redefinition of
$\alpha_1, \alpha_3$ : $ \alpha_1 \to \alpha_1 + \frac{\epsilon_0 r_0
}{r_1},  \alpha_3 \to \alpha_3 - \frac{\epsilon_0 r_0
}{r_1}$. 
After this redefinition the solution \rf{C=monehalfsols} describes the
two configurations with a common $q=r_1/2$ and $\alpha_i$'s given by $\alpha_1=-\alpha_3,
\alpha_2=1/2$ and $\alpha_1=\alpha_3, \alpha_2=1/2-\alpha_1$. In
\cite{gromak-book} the  B\"acklund transformations $s_i,i=0,1,2,3$ 
were used to obtain from this solution
the rational solutions other than the solutions to 
the Riccati equations.

\subsubsection{$\mathbf{C=0}$}
Plugging the ansatz \rf{pansatz} and expression \rf{qansatz} into equations
\rf{qz} and \rf{pz}
we find that for $C=0$ there are only two allowed values for $p_0$: 
\begin{equation}
p_0 = 0 \quad \text{or} \quad p_0= \frac{\alpha_1-\alpha_3}{r_1} \, ,
\lab{p0cases}
\end{equation}
while the allowed values for $p_1$ are 
\begin{equation}
p_1 = 0 \quad \text{or} \quad p_1= -r_0 \quad \text{or} \quad 
p_1= \frac{1}{2} r_0 \left(\frac{4 \epsilon _0}{r_1^2}-1\right)\, .
\lab{p1values}
\end{equation}
Given that the $\pi^2$ symmetry \rf{pisquare}
transforms solutions with $p_0=0, p_1=-r_0$ and $p_0=0, p_1=0$ into each other we will only need
to describe two of the three  cases shown in \rf{p1values} for $p_0=0$.
Discarding those solutions that are connected with divergences in
\rf{qansatz} (i.e. Riccati solutions with $p_0=p_1=0$ or $p_0=0,p_1=-r_0$) leaves  us with a number of solutions that can be parametrized
by one or two variables.

There are two choices for the value of $p_0$ according  to equation
\rf{p0cases}.
Considering the only  remaining case of $p_0=0$ 
and  $p_1\ne 0, -r_0$ from equation \rf{p1values}
we obtain two possible solutions :
\begin{align}
p_0&=0, \quad p_1= \frac{1}{2} r_0 \left(\frac{4 \epsilon
_0}{r_1^2}-1\right),\qquad\alpha_1=\alpha_3 \nonumber\\
\alpha_2&= -\frac{\left(2 \alpha_3-1\right) \left(r_1^2-4 \epsilon
_0\right)}{2 r_1^2},
\quad q (z) =\frac{r_1}{2}
\lab{qr1over2}
\end{align}
and 
\begin{align}
p_0&=0, \quad p_1= \frac{1}{2} r_0 \left(\frac{4 \epsilon _0}{r_1^2}-1\right),\qquad
\alpha_1=\alpha_3 \nonumber\\
\alpha_2&= \frac{\left(2 \alpha_3-1\right) \left(r_1^2-4
\epsilon_0\right)}{8 \epsilon_0},
\quad q (z) =
r_1 \frac{(2\alpha_3-1)r_1+2zr_0\epsilon_0}{4zr_0\epsilon_0} \, ,
\lab{a1eq22}
\end{align}
with $\epsilon_0$ and $\alpha_3$ selected as two free parameters.

For $p_0= (\alpha_1-\alpha_3)/r_1\ne 0$ we  
find a class of %
solutions with 
\begin{equation}
\alpha_2
= \pm \frac{\left(r_1^2-4 \epsilon_0\right)}{2 r_1^2}\,.
\lab{C0a2}
\end{equation}
This class of solutions is defined by values of the sum
$\alpha_1+\alpha_3$ with four possibilities :
\[
\alpha_1+\alpha_3 = 1 \pm \frac{4 \epsilon_0}{r_1^2}, \qquad
\alpha_1+\alpha_3 = 1 \mp 1 \, .
\]
The plus/minus signs in the above equation are correlated with 
the plus/minus signs in expression \rf{C0a2} for $\alpha_2$.

We first present solutions for the cases $\alpha_1+\alpha_3 = 1 \pm 
4 \epsilon_0/r_1^2$ :
 \begin{align}
 p_1&= \frac{1}{2} r_0 \left(\frac{4 \epsilon _0}{r_1^2}-1\right),
\qquad \alpha_2= \pm \frac{\left(r_1^2-4 \epsilon_0\right)}{2 r_1^2},
\nonumber \\
\alpha_1&=-\alpha_3 +1 \pm \frac{4 \epsilon_0}{r_1^2},
\quad  p_0=-\frac{\mp4\epsilon_0+r_1^2(2\alpha_3-1)}{r_1^3}
\nonumber  \\
q (z) &=\frac{r_1}{2}\frac{r_0r_1 z (r_1^2 \pm 4\epsilon_0)
+4r_1^2( \pm1 \mp \alpha_3)+16\epsilon_0}{r_0r_1 z(r_1^2 \pm
4 \epsilon_0)+2r_1^2(\pm1 \mp 2 \alpha_3)+8\epsilon_0} \, .
 \lab{p0general}
\end{align}
Solutions for the remaining cases of $\alpha_1+\alpha_3 = 1 \pm 1$
are given by 
\begin{align}
 p_1&= \frac{1}{2} r_0 \left(\frac{4 \epsilon _0}{r_1^2}-1\right),
\qquad \alpha_2= \pm \frac{\left(r_1^2-4 \epsilon_0\right)}{2 r_1^2},
\nonumber \\
\alpha_{1 }&=-\alpha_3 +1 \mp 1,  
\nonumber\\
q_{-} (z) &=\frac{r_1 (z^2  r_0^2 (-16 \epsilon_0^2+r_1^4)-16 r_1^2 r_0 z p_0
\epsilon_0-4r_1^4 p_0^2-8r_1^3 p_0)}{2(zr_0r_1^2-4r_0z\epsilon_0-2r_1^2p_0)
(zr_0r_1^2+4r_0z\epsilon_0+2r_1^2p_0)}
\nonumber\\
q_{+} (z) &= \frac{r_1}{2} \, ,
\lab{strange} 
\end{align}
with the condition $p_0=(\alpha_1-\alpha_3)/r_1$ being satisfied.
Note that $\lim_{z \to \infty} q = r_1/2$ for all the above results
in \rf{qr1over2}, \rf{a1eq22}, \rf{p0general} and \rf{strange}
in agreement with one of the results of Subsection \ref{subsection:kitaev}.

For $p_1=0, p_0 \ne 0$ there is a one-parameter class of solutions with $\vert \alpha_2 \vert
=1$:
\[
\alpha_1-\alpha_3= p_0 r_1, \quad 
\alpha_1+\alpha_3= 1, \quad \alpha_2
= \pm 1 \, ,
\]
for which we obtain:
\begin{equation}
 p_0=X, \qquad q (z)=
\frac{r_0r_1zX(r_1^2-4\epsilon_0)+X(r_1^2-4\epsilon_0) \mp r_0z 
(r_1^2-4\epsilon_0)+r_1}{2(zr_0X(r_1^2-4\epsilon_0)+1)} \, ,
\lab{alpha2e1}
\end{equation}
where $X$ is a root of equation 
\[X^2( r_1^2-4\epsilon_0)-1=0\, .
\]
Furthermore there is also a two-parameter solution with $p_1=0$ and
$p_0=(\alpha_1-\alpha_3)/r_1\ne 0$ for which $\alpha_2$ and $\alpha_3$
are determined in terms of $\alpha_1, \epsilon_0$ from conditions:
\[
\begin{split}
\alpha_2 &= X \frac{\alpha_1-\alpha_3}{r_1} \\
0&=X\left( \alpha_1-\alpha_3+\alpha_3^2-\alpha_1^2 \right)/r_1 +
\alpha_1+\alpha_3-\alpha_3^2-\alpha_1^2 +2\epsilon_0 \left(
\alpha_1-\alpha_3\right)^2/r_1^2 \, ,
\end{split}
\]
where again $X$ is a root of equation $X^2( r_1^2-4\epsilon_0)-1=0$.
Corresponding expressions for solutions $q(z)$ are
\begin{equation}
q(z)= r_1 \left( r_0 z (r_1-X) +2 \alpha_1 \right) \left(
-r_1^2+2\epsilon_0+X r_1\right) D^{-1} \, ,
\lab{alpha2X}
\end{equation}
where the denominator $D$ is given by
\[
\begin{split}
D&= 2r_1\alpha_1(-r_1+X)-Xr_1+r_1^2\pm \sqrt{2}
\sqrt{-r_1^2(-r_1^2+2\epsilon_0+Xr_1)} \\
&-2  r_0r_1z (r_1^2-2\epsilon_0 +r_1X) \,.
\end{split}
\]
Limits for $z \to \infty$ for $q(z)$ given in \rf{alpha2e1} and \rf{alpha2X}
are $r_1/2+X/2$, in agreement with Kitaev et al type of expressions
from Subsection \ref{subsection:kitaev}.

There is also a constant solution with $p_1=0$ with $p_0$ and
$\epsilon_0$ being parameters :
\begin{equation*}
\begin{split}
\alpha_3 & =Yp_0 ,\alpha_{1}=Yp_0+r_1 p_0, \;\; 
\alpha_2=-\frac{p_0(-2Y-r_1+(r_1^2-4\epsilon_0)p_0)}{-1+2Y p_0+r_1p_0}, \\
q (z)&=
\frac{r_1(-1+r_1p_0)+r_1^2p_0-Y-2\epsilon_0p_0}{-1+2Yp_0+r_1p_0} \, ,
\end{split}
\end{equation*}
where $Y$ is a root of equation $Y^2+\epsilon_0+r_1Y=0$.

\subsubsection{$C=-1$}
Plugging the ansatz \rf{pansatz} and expression \rf{qansatz} into equations
\rf{qz} and \rf{pz}
we obtain for $C=-1$ that $ p_0=0$ and $p_1$ must take one of the following values:
\begin{equation*}
p_1=0,\; \text{or} \;  p_1=-r_0 ,\; \text{or} \;  p_1=-r_0/2 \, .
\end{equation*}
The first two solutions correspond to singularities in \rf{qansatz}
that are connected with Riccati cases.

We will now focus on the remaining case of  $ p_0=0$ and $p_1=-r_0/2$ for
$\epsilon_0 \ne 0$ that implies a relation:
\begin{equation*}
\alpha_1=\alpha_3 \, .
\end{equation*}
Plugging $p (z)=-r_0\, z/2 $ and $\alpha_1=\alpha_3$ into the underlying
equations we arrive at a single two-parameter solution :
\begin{equation}
q(z) = \frac12 \frac{ z r_1 (2\alpha_3 -1) +2 \epsilon_0 r_0}{z
(2\alpha_3 -1) } \, ,
\lab{single2para}
\end{equation}
in terms of two parameters $\alpha_3, \epsilon_0$ and with 
\begin{equation*}
\alpha _2= 
\frac{-\left(1-2 \alpha _3\right){}^2+r_0^2 \epsilon _0}
{2(2 \alpha _3-1)}\,.
\end{equation*}

\subsubsection{$C=1/2$}
For $C=1/2$ ($J=-2$) we find a generalization of formula  \rf{Cothersols} to
values of $\alpha_2$ other than  $\alpha_2=-1/2$ required by \rf{Cothersols}.
This generalized  solution is based on  $p(z)$
given by:
\begin{equation}
p(z)=p_0+p_1 z+p_2 z^2 = -\frac{r_1^2r_0 (4
\alpha_2^2-1)}{32\epsilon_0}-\frac{r_0}{2}\, z + \frac{2 r_0 \epsilon
_0}{r_1^2} \, z^2 \,,
\lab{pzChalf}
\end{equation}
that for $\alpha_2= \pm 1/2$ coincides with the
formula \rf{pgeneral}.

The two solutions corresponding to \rf{pzChalf} when parametrized by
$\alpha_2$ and $ \epsilon_0$ will have two possible choices for
the parameter $\alpha_1$ and the corresponding solutions $q(z)$:
\begin{equation}
\begin{split}
\alpha_1 &=\frac{32 \epsilon_0+r_1^3r_0(1-4 \alpha_2^2)}{64
\epsilon_0}\\
q (z) &=\frac{r_1 (64 z^2  r_0 \epsilon_0^2-32 z \epsilon_0 r_0 r_1^2
\alpha_2 -32 \epsilon_0 r_1 +4 r_0 r_1^4 \alpha_2^2- r_0 r_1^4)}{2r_0 (8z\epsilon_0 -2 r_1^2
\alpha_2 -r_1^2) (8z\epsilon_0 -2 r_1^2
\alpha_2 +r_1^2)} \, ,
\end{split}
\lab{alpha1Chalf1}
\end{equation}
and 
\begin{equation}
\begin{split}
\alpha_1 &=\frac{32 \epsilon_0 (1-2\alpha_2)+r_1^3r_0(1-4
 \alpha_2^2)}{64 \epsilon_0} \\
 q (z) &=\frac{r_1 (64 z^2  r_0 \epsilon_0^2-32 \epsilon_0 r_1 (1+2
\alpha_2) -4 r_0 r_1^4 \alpha_2 (1 + \alpha_2) - r_0 r_1^4)}{2r_0 (8z\epsilon_0 -2 r_1^2
\alpha_2 -r_1^2) (8z\epsilon_0 +2 r_1^2
\alpha_2 +r_1^2)} \, ,
\end{split} 
\lab{alpha1Chalf2}
\end{equation}
with $\alpha_3$ such that $\alpha_3= \alpha_1-p_0 r_1$ with $p_0$ from
eq.\rf {pzChalf}.

\subsection{Solutions with two non-zero deformation parameters
$\epsilon_0,\epsilon_1$}
\label{subsection:two}
In this case the mixed Painlev\'e P$_{III-V}$ equations have only
invariance under the automorphism $\pi$ from \rf{ext-pi}.
The ansatz $p(z)=p_0+p_1 z$ leads to rational solutions for $q(z)$
even when both deformation parameters are present. 
In this case we encounter three-parameter solutions 
for two values of $C$ parameter:  $C=0$ and $C=-1$.
\subsubsection{The case of $\mathbf{C=0}$}
For $C=0$ the only solution with $q(z) = r_1/2$ present in this 
case is conveniently expressed in terms of
$p_1,\alpha_1,\alpha_3$ as follows:
\begin{equation*}
\begin{split}
p_0&=\frac{\alpha_1-\alpha_3}{r_1}, \quad \alpha_2=
\frac{p_1(\alpha_1+\alpha_3-1)}{r_0} \\
\epsilon_0&= \frac{r_1^2(r_0+2p_1)}{4r_0} , \quad 
\epsilon_1= \frac{\alpha_1^2-\alpha_3^2}{r_1^2}\, .
\end{split}
\end{equation*}
\subsubsection{The case of $\mathbf{C=-1}$}
For $C=-1$ there are three  solutions, each with $p_0=0$,
and each expressed 
by three-parameters here chosen to 
$\alpha_1,\alpha_2, \alpha_3$.
First there is a solution with $\alpha_1 \ne \alpha_3$: 
\begin{equation*}
q(z)= \frac{(\alpha_3+\alpha_1+2 \alpha_2-1)(\alpha_1+\alpha_3) 
+r_0 z r_1 \alpha_1}{r_0 z (\alpha_3+\alpha_1)}
\end{equation*}
and
\begin{equation*}
\begin{split}
p_0&=0, \quad p_1=-r_0/2 \\
\epsilon_0&=
\frac{(\alpha_1+\alpha_3-1)(\alpha_1+2\alpha_2+\alpha_3-1)}{r_0^2}
, \quad 
\epsilon_1= \frac{(\alpha_1-\alpha_3)r_0^2}{4(\alpha_1+\alpha_3)} \, .
\end{split}
\end{equation*}
The two remaining solutions are given in terms  
$\epsilon_0,\alpha_1,\alpha_3$ and map into
each other under the substitution 
$\alpha_1 \leftrightarrow \alpha_3, \epsilon_1 \to -
\epsilon_1$. We give expressions for one of the solutions:
\begin{equation*}
\begin{split}
p_0&=0, \quad p_1=\frac{-\epsilon_0
r_0^2+\alpha_1\alpha_3-\alpha_1}{2\epsilon_0 r_0} \\
\alpha_2&=
\frac{(-\epsilon_0
r_0^2+\alpha_1\alpha_3-\alpha_1)(-\epsilon_0
r_0^2-2\alpha_3+1+\alpha_3^2)}{2 \epsilon_0 r_0^2(\alpha_3-1)} \\
\epsilon_1&= \frac{(-\epsilon_0
r_0^2+\alpha_1\alpha_3-\alpha_1)(\epsilon_0
r_0^2+\alpha_1\alpha_3-\alpha_1)}{4 \epsilon_0^2 r_0^2}
\end{split}
\end{equation*}
and with
\[
q(z)= \frac{r_0\epsilon_0}{z(\alpha_3-1)}\, .
\]
For the other solution obtained by substitution 
$\alpha_1 \leftrightarrow \alpha_3, \epsilon_1 \to -
\epsilon_1$ applied on $p_1,\alpha_2, \epsilon_1$
it holds that 
\[
q(z)= \frac{r_0\epsilon_0}{z(\alpha_1-1)} +r_1 \,.
\]

\section{Discussion }
\label{sec:discussion}

Having formulated the  mixed Painlev\'e  III - V  model and
its symmetries for one or two non-zero deformation parameters in \cite{AGZ2016}
we were interested in finding its solutions in anticipation that
these solutions will show features that will  reflect non-conventional
symmetries and structures of parameters of the underlying model.
Most of this
paper is focused on a search for rational solutions in the setting of
P$_{III-V}$  model with one non-zero deformation parameter
that could be chosen to be $\epsilon_0$  without any loss of generality.
Given a different symmetry of the underlying model for the non-zero
 $\epsilon_0$ as compared with the symmetry of  P$_{V}$ equation one would 
 expect 
 that the structure of rational solutions of the model
 would be vastly different from those obtained for the zero value of
 $\epsilon_0$. This is confirmed by results of this paper. 
 Dependence of solutions on  two (and even, in some cases, three) 
 independent parameters 
is clearly a feature  that can be directly related to  
 richer parameter space of the underlying equation. There are some
 features of solutions that were less expected.
For example, although the formulation of 
P$_{III-V}$  model with its Weyl symmetry group
$W[s_0,s_2, \pi^{2}]$ can be developed 
for an arbitrary value of the parameter $C$,
apart from two exceptions we were only able to obtain its rational solutions 
as well as solutions to the underlying Riccati equations
for two values of $C$, namely $C=0$ and $C=-1$.
The two exceptions to this rule are $C=-1/2$ (but that is not surprising given that in this case the
model is equivalent to P$_V$ equation and solutions to 
P$_V$ are well known
\cite{clarksonp5,clarkson2006,gromak-book,kitaev,masuda,mok}) 
and, more surprisingly,  
$C=1/2$ for which we were able to find rational solutions described by relations
\rf{pzChalf}, \rf{alpha1Chalf1} and \rf{alpha1Chalf2}. 

In analyzing the structure of solutions for different values of the
underlying parameters one needs to recognize 
limitations imposed by structure of the Weyl symmetry
group $W[s_0,s_2, \pi^{2}]$ of P$_{III-V}$. 
Contrary to the situation present in  P$_{III}$ and  P$_{V}$ models 
the underlying symmetry of P$_{III-V}$ does not allow 
for construction of  non-Riccati
solutions out of Riccati solutions. The reason is simply that $s_0$ and 
$s_2$ never change the underlying constraint $\alpha_0=0$. The
automorphism $\pi^2$ flips only the $\alpha_0$ and $\alpha_2$
parameters without essentially changing the nature of solution. 
Thus the Toda-type structure of B\"acklund transformations common in 
P$_{III}$ and  P$_{V}$ models is absent in P$_{III-V}$ and 
does not give rise to determinant solutions
build out of Riccati solutions. Although the Toda like symnmetry structure is evidently
missing from the P$_{III-V}$ model we remarkably find a class of rational
solutions of the underlying Riccati equation with parameters that are
``quantized'' by integer $n \in \mathbb{Z}$ (see equations
\rf{nepsilon0C0}-\rf{qC0nm2}).

While B\"acklund symmetry
of P$_{III-V}$ model does not offer ability to 
generate a chain of solutions of increased complexity which exists in
submodels the trade-off is provided by much richer parameter space. 
This is reflected in existence of two-parameter solutions of  P$_{III-V}$ which can be reduced
to solutions of submodels by fixing one of these parameters to 
special values. We have provided explicit examples (see below equations \rf{qC0n1}
and \rf{qC0nm2}) of how by projecting on special directions in the
parameter space of P$_{III-V}$ model one recovers solutions
P$_{V}$ and P$_{III}$ from solutions of P$_{III-V}$ 
with parameters that are ``quantized'' by integer 
$n \in \mathbb{Z}$.
This understanding also provides a heuristic
argument for why solutions P$_{III-V}$ can not be recovered from 
solutions to P$_{V}$ or P$_{III}$ by simple rescaling of parameters
of P$_{V}$ or P$_{III}$. 
Efforts to better understand algebraic relations between solutions 
of different Painlev\'e equations are an active area of research 
and recently progress has been made that imply
that there are no general B\"acklund
transformations between generic Painlev\'e 
equations from different families P$_{I}$-P$_{V}$ \cite{nagloo}. It
should be interesting to include P$_{III-V}$ model in such studies.

Let us also discuss potential 
limitations of the strategy used to obtain rational  solutions by
computer algebraic methods.
One such potential limitation was a key assumption of $p(z)$ being 
a finite Laurent series.. It will be interesting to explore if one can lift such
condition and what it would imply for possible additional solutions.

\begin{appendices}
\section{Alternative Solutions of Riccati Equations in terms of
Laguerre Functions}
\label{sec:alternativeRiccati}

A linearized version
of the Riccati equation \rf{lricattieq} has been solved in this paper
in terms Whittaker functions for three values of $J=-1,0,1$.
In this Appendix we provide alernative solution for  \rf{lricattieq}
with the same values of $J$ as linear combination of 
of the generalized Laguerre functions
$ \text{L}[\nu,\lambda,z]$ and the
confluent hypergeometric functions $\text{U}[\nu, \lambda,z]$ \cite{arfken,weisstein}:
\begin{align}
w_{C=0}&=e^{\frac{1}{2} z r_0 \left(r_1-\sqrt{r_1^2+4 \epsilon _0}\right)} 
\Big( c_1 \text{U}\left[\frac{\alpha_1+\alpha _3}{2}
+\frac{r_1(\alpha_1-\alpha_3)}{2\sqrt{r_1^2+4 \epsilon _0}},\alpha _1+\alpha _3,z r_0
\sqrt{r_1^2+4 \epsilon _0}\right]
\nonumber
\\
&+c_2 \text{L}\left[
-\frac{\alpha_1+\alpha _3}{2}
-\frac{r_1(\alpha_1-\alpha_3)}{2\sqrt{r_1^2+4 \epsilon _0}},\alpha_1+\alpha_3-1,z r_0
\sqrt{r_1^2+4 \epsilon _0}\right]\Big)
\lab{wc0laguerre}
\end{align}
\begin{equation}
w_{C=-\frac{1}{2}}=c_1 \text{U}\left[\alpha _1+\frac{r_0 \epsilon _0}{r_1},
\alpha _1+\alpha _3,z r_0 r_1\right]+ c_2 \text{L}\left[-\alpha _1
-\frac{r_0 \epsilon _0}{r_1},\alpha _1+\alpha _3-1,z r_0 r_1\right]
\lab{wcmhalflaguerre}
\end{equation}
\begin{align}
\omega_{C=-1}&=z^{\frac{1}{2} \left(1-\alpha _1-\alpha _3+\xi\right)}
\Big(c_1\text{U}\left[\frac{1}{2} \left(1+\alpha _1-\alpha _3+\xi\right),1
+\xi,z r_0 r_1\right] \nonumber\\
&+c_2 \text{L}\left[\frac{1}{2} \left(-1-\alpha _1+\alpha _3-\xi\right),
\xi,z r_0 r_1\right]\Big) \,, 
\lab{wcmonelaguerre}
\end{align}
where $\xi=\sqrt{(\alpha _1+\alpha _3-1)^2+4 r_0^2 \epsilon _0}$.

The above solutions \rf{wcmhalflaguerre} for $C=-1/2$ appeared in the context of
P$_V$ model \cite{clarksonp5,clarkson2006,masuda,mok}	

Our wish is to be able to express the hypergeometric functions,
solutions of the $P_{III-V}$ associated Ricatti equation, in a
systematic way as ratio of polynomials.

Dealing only with the hypergeometric part of the solution, we 
are able to do so.

\subsection{C=0}
\label{subsec:C0}
When we solve the above equation \rf{ricattieq}  for $C=0$ and 
impose the quantized values ($n\in \mathbb{Z}$):
\begin{equation*}
\epsilon _0\to -\frac{\left(n+\alpha_1\right) \left(n+\alpha_3\right)
r_1^2}{\left(2 n+\alpha_1+\alpha_3\right){}^2} \, ,
\end{equation*}
we obtain:
\begin{equation}
q(z)=\frac{r_1 \left(H_1-H_2 \alpha _1+H_2 \alpha
_3\right)-\sqrt{\frac{r_1^2 \left(\alpha _1-\alpha
_3\right){}^2}{\left(2 n+\alpha _1+\alpha _3\right){}^2}}
\left(H_1+H_2 \alpha _1+H_2 \alpha _3\right)}{2 H_1}\,,
\label{75}
\end{equation}
where
\begin{align}
H_1=&\text{U}\left[\frac{r_1 \left(\alpha _1-\alpha _3\right)+\left(\alpha _1+\alpha _3\right) \sqrt{\frac{r_1^2 \left(\alpha _1-\alpha _3\right){}^2}{\left(2 n+\alpha _1+\alpha _3\right){}^2}}}{2 \sqrt{\frac{r_1^2 \left(\alpha _1-\alpha _3\right){}^2}{\left(2 n+\alpha _1+\alpha _3\right){}^2}}},\alpha _1+\alpha _3,z r_0 \sqrt{\frac{r_1^2 \left(\alpha _1-\alpha _3\right){}^2}{\left(2 n+\alpha _1+\alpha _3\right){}^2}}\right]\\
H_2=&\text{U}\left[\frac{r_1 \left(\alpha _1-\alpha
_3\right)+\left(2+\alpha _1+\alpha _3\right) \sqrt{\frac{r_1^2
\left(\alpha _1-\alpha _3\right){}^2}{\left(2 n+\alpha _1+\alpha
_3\right){}^2}}}{2 \sqrt{\frac{r_1^2 \left(\alpha _1-\alpha
_3\right){}^2}{\left(2 n+\alpha _1+\alpha _3\right){}^2}}},1+\alpha _1+\alpha _3,z r_0 \sqrt{\frac{r_1^2 \left(\alpha _1-\alpha _3\right){}^2}{\left(2 n+\alpha _1+\alpha _3\right){}^2}}\right] \, .
\end{align}
Although the above expressions are solutions for both positive or the
negative values of the square root, the confluent hypergeometric
function $U$ assumes polynomial values just for negative integers of
its first argument.  For that reason we choose the sign of the square
root depending on the sign of $n$. Consistently for 
negative values  of $n$ we choose the negative sign for the root, 
getting from (\ref{75}):
\[
\begin{split}
q(z)&=\frac{r_1 \left(-\alpha _1^2-\alpha _1 n+\alpha _3 \left(\alpha _3+n\right)\right) 
U\left(n+\alpha _1+\alpha _3+1,\alpha _1+\alpha _3+1,\frac{z r_0 r_1
\left(\alpha _1-\alpha _3\right)}{2 n+\alpha _1+\alpha
_3}\right)}{\left(\alpha _1+\alpha _3+2 n\right) U\left(n+\alpha _1+\alpha _3,\alpha _1+\alpha _3,\frac{z r_0 r_1 \left(\alpha _1-\alpha _3\right)}{2 n+\alpha _1+\alpha _3}\right)}\\
&+\frac{r_1 \left(\alpha _3+n\right)}{\alpha _1+\alpha _3+2 n} \, ,
\end{split}
\]
which are polynomials (although more compactly written as 
$U$ functions) due to identities %
\begin{align}
U[a,b,z]=&z^{1-b}U[a-b+1,2-b,z]\\
U[-n,b,z]=&(-1)^nn! L_n^{b-1}(z) \, .
\end{align}
The first identity shows that $U$ appearing in the above expression
for $q(z)$ can be rewritten to have the first argument of $U$ equal 
to $n+1$. The second identity tells us that it 
becomes a Laguerre polynomial, once $n$ is a negative integer.

For non-negative values  of $n$ we must choose the 
positive sign for the root, getting:
\begin{equation}
q(z)=\frac{\left(\alpha _3-\alpha _1\right) n r_1 U\left(1-n,\alpha
_1+\alpha _3+1,-\frac{z r_0 r_1 \left(\alpha _1-\alpha _3\right)}{2
n+\alpha _1+\alpha _3}\right)}{\left(\alpha _1+\alpha _3+2 n\right)
U\left(-n,\alpha _1+\alpha _3,-\frac{z r_0 r_1 \left(\alpha _1-\alpha
_3\right)}{2 n+\alpha _1+\alpha _3}\right)}+\frac{r_1 \left(\alpha _1+n\right)}{\alpha _1+\alpha _3+2 n} \, .
\end{equation}

\subsection{C=-1}
\label{subsec:Cm1}
In this case we insert   :
\begin{equation*}
\epsilon _0\to \frac{\left(n+\alpha _1\right) \left(1+n-\alpha _3\right)}{r_0^2}
\end{equation*}
in (\ref{75}) leading to solution:
\begin{equation}
q(z)=\frac{-\left(H_3+z H_4 r_0 r_1\right) \alpha _1+\left(H_3-z H_4
r_0 r_1\right) \left(1+\sqrt{\left(1+2 n+\alpha _1-\alpha
_3\right){}^2}-\alpha _3\right)}{2 z H_3 r_0} \, ,
\end{equation}
where
\begin{align}
H_3=&U\left(\frac{1}{2} \left(\alpha _1-\alpha _3+\sqrt{\left(\alpha _1-\alpha _3+2 n+1\right){}^2}+1\right),\sqrt{\left(\alpha _1-\alpha _3+2 n+1\right){}^2}+1,r_0 r_1 z\right)\\
H_4=&U\left(\frac{1}{2} \left(\alpha _1-\alpha _3+\sqrt{\left(\alpha
_1-\alpha _3+2 n+1\right){}^2}+3\right),\sqrt{\left(\alpha _1-\alpha
_3+2 n+1\right){}^2}+2,r_0 r_1 z\right) \, .
\end{align}
As before, the choice of signs is determined by the value of $n$. 
For negative values of $n$ we choose the positive root resulting 
in:
\begin{equation}
q(z)= \frac{-a_3+n+1}{r_0 z}-\frac{r_1 \left(a_1-a_3+n+1\right)
U\left(n+a_1-a_3+2,2 n+a_1-a_3+3,z r_0
r_1\right)}{U\left(n+a_1-a_3+1,2 n+a_1-a_3+2,z r_0 r_1\right)} \,.
\end{equation}
For non-negative values of $n$ we choose a negative root 
resulting in:
\begin{equation}
q(z)= \frac{n r_1 U\left(1-n,-2 n-a_1+a_3+1,z r_0
r_1\right)}{U\left(-n,-2 n-a_1+a_3,z r_0 r_1\right)}-\frac{a_1+n}{r_0
z} \, .
\end{equation}

\end{appendices}

\subsection*{Acknowledgements}
  JFG and AHZ thank CNPq for financial support. VCCA acknowledges the
  research support of the  S\~{a}o Paulo Research Foundation (FAPESP)
  via grant number 2016/22122-9.

\end{document}